\documentclass[aps,prb,twocolumn,superscriptaddress]{revtex4}
\usepackage{graphicx}
\begin{document}

% Use the \preprint command to place your local institutional report
% number in the upper righthand corner of the title page in preprint mode.
% Multiple \preprint commands are allowed.
% Use the 'preprintnumbers' class option to override journal defaults
% to display numbers if necessary
%\preprint{}
%Title of paper
\title{Electron doping evolution of the magnetic excitations in BaFe$_{2-x}$Ni$_{x}$As$_{2}$}

\author{Huiqian Luo}
\thanks{These authors made equal contributions to this paper}
\affiliation{Beijing National Laboratory for Condensed Matter
Physics, Institute of Physics, Chinese Academy of Sciences, Beijing
100190, China}
\author{Xingye Lu}
\thanks{These authors made equal contributions to this paper}
\affiliation{Beijing National Laboratory for Condensed Matter
Physics, Institute of Physics, Chinese Academy of Sciences, Beijing
100190, China}
\author{Rui Zhang}
\affiliation{Beijing National Laboratory for Condensed Matter
Physics, Institute of Physics, Chinese Academy of Sciences, Beijing
100190, China}
\author{Meng Wang}
\affiliation{Beijing National Laboratory for Condensed Matter
Physics, Institute of Physics, Chinese Academy of Sciences, Beijing
100190, China}
\author{E. A. Goremychkin}
\affiliation{ISIS Facility, Rutherford Appleton Laboratory, Chilton, Didcot, Oxfordshire OX11 0QX, UK}
\author{D. T. Adroja}
\affiliation{ISIS Facility, Rutherford Appleton Laboratory, Chilton,
Didcot, Oxfordshire OX11 0QX, UK}
\author{Sergey Danilkin}
\affiliation{Bragg Institute, Australian Nuclear Science and
Technology Organization, New Illawarra Road, Lucas Heights NSW-2234
Australia}
\author{Guochu Deng}
\affiliation{Bragg Institute, Australian Nuclear Science and
Technology Organization, New Illawarra Road, Lucas Heights NSW-2234
Australia}
\author{Zahra Yamani}
\affiliation{Canadian Neutron Beam Centre, National Research
Council, Chalk River Laboratories, Chalk River, Ontario K0J 1J0,
Canada}
\author{Pengcheng Dai}
%\footnote[2]{Electronic address of the corresponding author: pdai@rice.edu}
\email{pdai@rice.edu}
\affiliation{Department of Physics and Astronomy, Rice University, Houston, Texas 77005, USA}
\affiliation{Beijing National Laboratory for Condensed Matter Physics, Institute of Physics,
Chinese Academy of Sciences, Beijing 100190, China}


%\maketitle must follow title, authors, abstract, \pacs, and \keywords
\begin{abstract}
We use inelastic neutron scattering (INS) spectroscopy to study the
magnetic excitations spectra throughout the Brioullion zone
in electron-doped iron pnictide superconductors BaFe$_{2-x}$Ni$_{x}$As$_{2}$ with $x=0.096,0.15,0.18$.
While the $x=0.096$ sample is near optimal superconductivity with $T_c=20$ K and has
coexisting static incommensurate magnetic order,
the $x=0.15,0.18$ samples are electron-overdoped with reduced
$T_c$ of 14 K and 8 K, respectively, and have no static antiferromagnetic (AF) order.
In previous INS work on undoped ($x=0$) and electron optimally doped ($x=0.1$) samples,
the effect of electron-doping was found to
modify spin waves in the parent compound BaFe$_2$As$_2$
below $\sim$100 meV and induce a neutron spin resonance at the commensurate
AF ordering wave vector that couples with superconductivity.
While the new data collected on the $x=0.096$ sample
confirms the overall features of the
earlier work, our careful temperature dependent study of the resonance reveals that
the resonance suddenly changes its $Q$-width below $T_c$ similar to that of the optimally hole-doped
iron pnictides Ba$_{0.67}$K$_{0.33}$Fe$_2$As$_2$.  In addition, we establish the dispersion of the resonance and find it to change from
commensurate to transversely incommensurate with increasing energy.
Upon further electron-doping to overdoped iron pnictides with $x=0.15$ and 0.18,
the resonance becomes weaker and transversely incommensurate at all energies, while spin excitations above $\sim$100 meV
are still not much affected.  Our absolute spin excitation intensity measurements throughout the Brillouin zone for $x=0.096,0.15,0.18$
confirm the notion that the low-energy spin excitation coupling with itinerant electron is important for superconductivity in these materials, even though
the high-energy spin excitations are weakly doping dependent.
\end{abstract}

\maketitle

\section{Introduction}
Understanding the origin of superconductivity in strongly correlated electron materials is at the
forefront of modern condensed matter physics \cite{PALee,scalapino}.  Since high-transition temperature (high-$T_c$) superconductors such as
copper oxides and iron pnictides are derived from electron or hole-doping to their antiferromagnetic (AF) order parent compounds
\cite{PALee,scalapino}, much efforts over the past 27 years have been focused on determining the role of short-range
spin excitations in the superconductivity of these materials \cite{fujita,dai,tranquada}.
From inelastic neutron scattering (INS) experiments on copper oxide superconductors, it is well established that the low-energy ($E\leq 100$ meV)
spin excitations persist throughout the doping-induced superconductivity dome and vanish
when superconductivity ceases to exist in the overdoped regime \cite{fujita}.
While INS failed to detect high-energy spin excitations near the AF ordering wave vector $(0.5,0.5)$ in overdoped copper oxides \cite{fujita},
recent resonant inelastic X-ray scattering experiments find that high-energy ($\geq 100$ meV) spin excitations
in reciprocal space near the origin are almost independent of hole-doping across the superconductivity dome \cite{dean,Tacon}.  In the case of iron pnictides (Fig. 1) \cite{kamihara,Rotter,assefat,ljli,cruz,jzhao,qhuang}, INS experiments on single crystals of electron-doped
BaFe$_{2-x}T_x$As$_2$ (where $T=$Co, Ni)
have mapped out the doping evolution of the spin excitations \cite{mdlumsden,sxchi,sli09,dkpratt09,adChristianson,dsinosov09,mywang10,clester10,jtpark10,hfli10,dsinosov11,mwang11,lharriger,kmatan,msliu12,hqluo12,gstucker12,mgkim13,mwang13}.
In the undoped state, BaFe$_2$As$_2$ exhibits nearly simultaneous
tetragonal-to-orthorhombic lattice distortion and collinear AF order below $T_N\approx 138$ K
[see left inset of Fig. 1(a)] \cite{qhuang}.  In the AF ordered state, BaFe$_2$As$_2$ forms randomly distributed
orthorhombic twin domains rotated 90$^\circ$ apart. As a consequence, the low-energy spin waves from the
two separate domains are centered around
the AF ordering wave vectors ${\bf Q_{AF}}=(\pm1,0)$ and $(0,\pm1)$, respectively, in reciprocal space [see right inset in Fig. 1(a)].
INS experiments using time-of-flight (TOF) chopper spectrometer at the ISIS spallation neutron source in UK
have measured spin waves of BaFe$_2$As$_2$ in absolute units
throughout the Brillouin zone and determined the spin-wave dispersions along the
two high symmetry directions as shown in the solid lines of Figs. 1(e) and 1(f) \cite{lharriger}.

\begin{figure*}[t]
\includegraphics[width=0.8\textwidth]{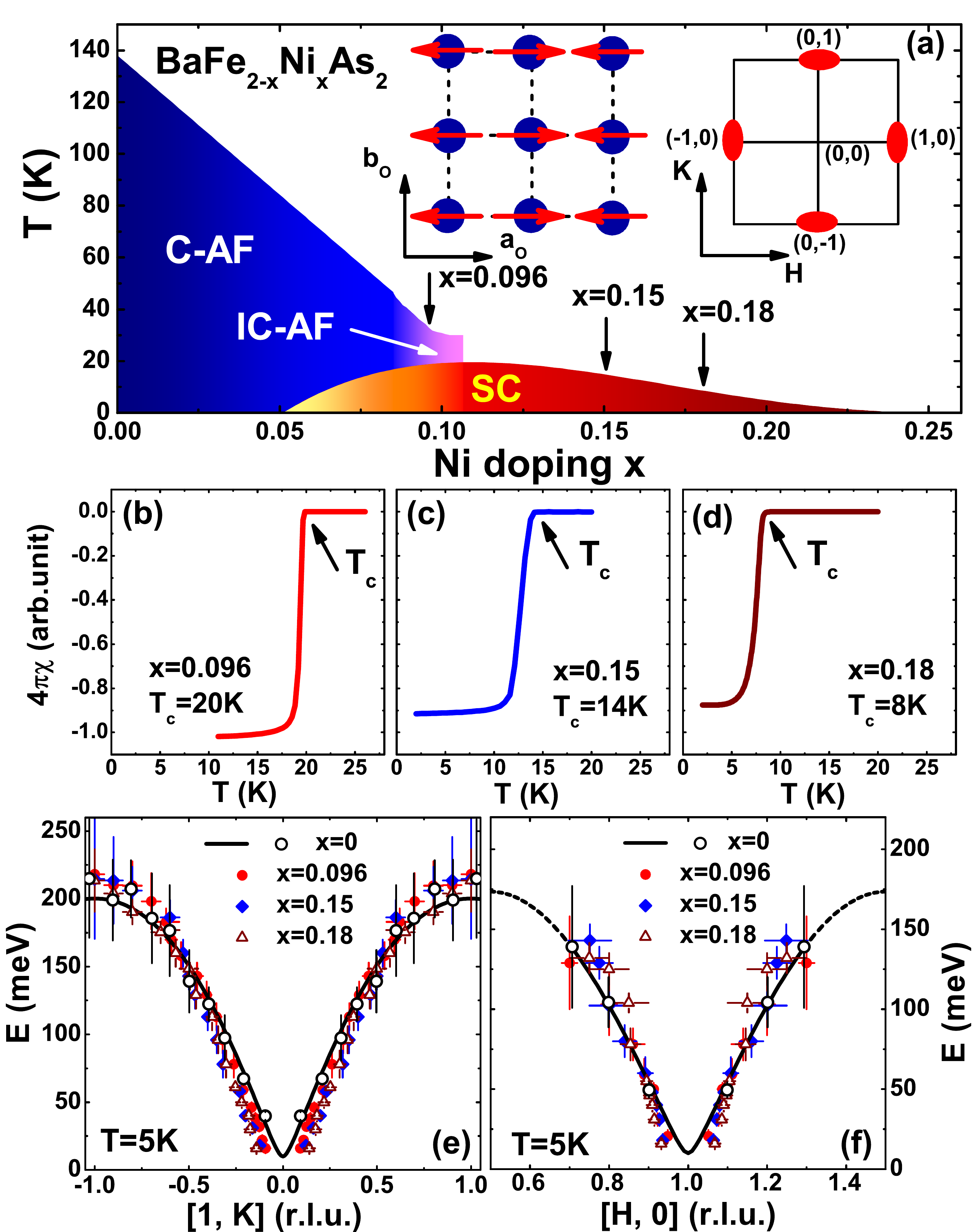}
\caption{(Color online)
(a) The schematic electronic phase diagram of
BaFe$_{2-x}$Ni$_{x}$As$_{2}$, where the arrows at $x=0.096,0.15,0.18$ indicate
doping levels studied in this paper \cite{hqluo,xylu13}. The Inserts show the in-plane magnetic structure in real space and Brillouin zone in reciprocal space.
(b,c,d) DC magnetic susceptibility indicates nearly 100\% diamagnetic volume for all three measured dopings with $T_c=20$ K, 14 K and 8 K.
(e,f) The dispersions of spin excitations along the
$[1,K]$ and $[H,0]$ directions for BaFe$_{2-x}$Ni$_{x}$As$_{2}$ with $x=0.096,0.15,0.18$.
The solid and dash lines are spin wave dispersions in the parent compound BaFe$_{2}$As$_{2}$ ($x=$ 0) \cite{lharriger}.
}
\end{figure*}

\begin{figure}[t]
\includegraphics[width=0.5\textwidth]{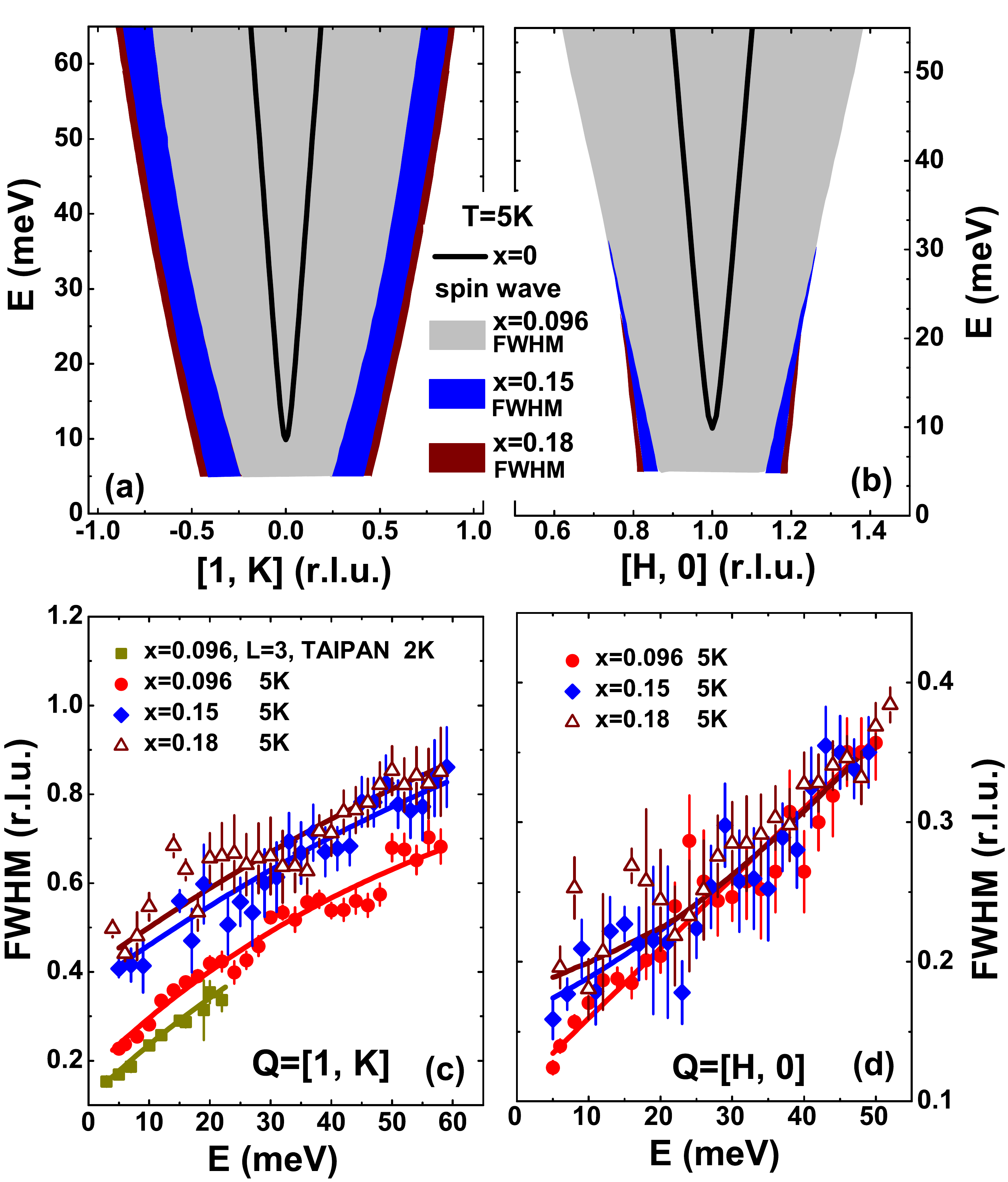}
\caption{(Color online)
Comparison of the dispersions of low-energy spin waves in BaFe$_2$As$_2$ with the FWHM of spin excitations in
BaFe$_{2-x}$Ni$_{x}$As$_{2}$ with $x=0.096,0.15,0.18$ along the $[1,K]$ and $[H,0]$ directions.
(a,b) The solid lines show spin wave dispersions of BaFe$_{2}$As$_{2}$ \cite{lharriger}.
The grey, blue, and brown regions show the FHWM of low-energy spin excitations of
BaFe$_{2-x}$Ni$_{x}$As$_{2}$ with $x=0.096,0.15,0.18$ along the $[1,K]$ and $[H,0]$ directions, respectively.
(c,d) Energy dependence of the FWHM along the transverse $[1,K]$ and longitudinal
$[H,0]$ directions as determined from TOF INS measurements.
The FWHM of peak along the $[1, K,3]$ direction from the
triple-axis experiments is also shown in (c).
 }
 \end{figure}

Figure 1(a) shows the schematic phase diagram of electron-doped BaFe$_{2-x}$Ni$_x$As$_2$ as determined from transport and
neutron diffraction experiments \cite{hqluo,xylu13}.  In previous work \cite{hqluo12}, the evolution of the low-energy spin excitations was found
to qualitatively follow the Fermi surface nesting picture and arise from quasiparticle excitations between
the hole and electron Fermi pockets near $\Gamma$ and $M$ points, respectively \cite{mazin2011n,Hirschfeld,Chubukov,kuroki,fwang}.
By comparing spin waves of the parent compound
with spin excitations of the
optimally electron-doped superconductor BaFe$_{1.9}$Ni$_{0.1}$As$_2$ in absolute units, it was found
that electron doping on BaFe$_2$As$_2$ affects only the low-energy spin excitations by broadening
the spin waves below 80 meV and forming a low-energy ($E_r\approx 7$ meV) neutron spin resonance below $T_c$,
but has no impact on spin waves above 100 meV \cite{msliu12}.  From systematic
triple-axis INS \cite{kmatan} and nuclear magnetic resonance \cite{flning10} measurements of the low-energy spin excitations in
 BaFe$_{2-x}$Co$_x$As$_2$, the suppression of superconductivity in electron-overdoped
 BaFe$_{2-x}T_x$As$_2$ is found to be associated with vanishing low-energy spin excitations.
Although these results are consistent with the presence of a large spin gap ($\sim$50 meV) in the electron-overdoped nonsuperconducting
BaFe$_{1.7}$Ni$_{0.3}$As$_2$ \cite{mwang13}, it is still unclear how spin excitations gradually
evolve from optimally doped superconductor to electron-overdoped nonsuperconductor.  Since spin excitations
may mediate electron pairing for superconductivity \cite{scalapino}, it would be important to determine the temperature
and electron-doping evolution of spin excitations in BaFe$_{2-x}$Ni$_x$As$_2$ across the superconductivity dome [Fig. 1(a)].

In this article, we report triple-axis and TOF INS studies of temperature and doping dependence of spin excitations
in BaFe$_{2-x}$Ni$_x$As$_2$.
For this work, we chose Ni-doping concentrations of $x=0.096,0.15$, and 0.18 with superconducting
transition temperatures of $T_c=20$ K [Fig. 1(b)], 14 K [Fig. 1(c)], and 8 K [Fig. 1(d)], respectively.
This range of Ni-dopings covers the nearly optimally electron-doped to electron-overdoped iron pnictide superconductors, and complements the earlier
work on the electron optimal $x=0.1$ superconductor \cite{msliu12} and the $x=0.3$ electron-overdoped nonsuperconductor \cite{mwang13}.
Consistent with earlier work \cite{clester10,jtpark10,hfli10,msliu12,hqluo12},
we find that the low energy spin excitations in BaFe$_{2-x}$Ni$_{x}$As$_{2}$ are transversely elongated ellipses around the
commensurate AF order wave vector.
 For the $x=0.096$ sample near optimal superconductivity,
a neutron spin resonance appears at $E_r=7$ meV below $T_c$, and the mode forms transversely incommensurate spin excitations at higher energies.
While the energy of the resonance is weakly temperature dependent,
the transverse and radial widths of the mode show a superconductivity-induced narrowing below $T_c$.
For samples at the overdoped side $x=0.15$, superconductivity induces a transversely incommensurate
resonance at $E_r=6.5$ meV.  On increasing electron-doping further to $x=0.18$, low-energy spin excitations have a broad commensurate
component independent of superconductivity and a transversely incommensurate resonance below $T_c$ at $E_r=5.5$ meV.
By comparing TOF INS data in
BaFe$_{2-x}$Ni$_{x}$As$_{2}$ with $x=0.096,0.15$, and 0.18, we establish the
wave vector and energy dependence of the
spin excitations throughout the Brillouin zone from optimally electron-doped to electron over-doped iron pnictides.
Our results are consistent with the idea that superconductivity in iron pnictides requires the
low-energy spin excitation-itinerant electron interaction \cite{mwang13}, and
indicate an intimate connection between spin excitations and superconductivity.

\section{Experiment}

We carried out INS experiments using the MERLIN TOF chopper spectrometer at the Rutherford-Appleton Laboratory, UK.
For the experiments, sizable single crystals of BaFe$_{2-x}$Ni$_{x}$As$_{2}$ grown by self-flux method \cite{ychen} were co-aligned on several aluminum plates by hydrogen-free glue
with both in-plane and out-of-plane mosaic less than 3$^\circ$.
The total mass of our samples is 41 grams for $x=0.096$, 45 grams for $x=0.15$,  25 grams for $x=0.18$, respectively.
Using orthorhombic crystalline lattice unit cell for easy comparison with the spin wave results of BaFe$_2$As$_2$ \cite{lharriger},
we define the wave vector $\bf Q$ at ($q_x$, $q_y$, $q_z$) as $(H,K,L) = (q_xa/2\pi, q_yb/2\pi, q_zc/2\pi)$ reciprocal lattice units (r.l.u.), where $a \approx b \approx 5.60$ \AA, and $c = 12.77$ \AA.  The samples are loaded inside a standard closed-cycle Helium refrigerator with incident beam parallel to the $c$-axis.
To probe spin excitations at different energies, we chose neutron incident beam energies of
$E_i= 20, 25, 30, 50, 80, 250, 450$ meV with
corresponding Fermi chopper frequencies of $\omega= 150, 200, 200, 400, 500, 550, 600$ Hz, respectively.
To facilitate comparison with spin waves in BaFe$_{2}$As$_{2}$ \cite{lharriger,harriger12},
spin excitations in doped materials are normalized to the absolute units (mbarn/sr/meV/f.u.) using a vanadium standard.
The neutron scattering cross section $S(Q,E)$ is related to the imaginary part of the
dynamic susceptibility $\chi^{\prime\prime}(Q,E)$ by correcting for the Bose population factor
via $S(Q,E)= 1/(1-\exp(-E/(k_BT)))\chi^{\prime\prime}(Q,E)$, where $k_B$ is the
Boltzmann's constant.  We can then calculate the local dynamic susceptibility by using
$\chi^{\prime\prime}(E)=\int{\chi^{\prime\prime}({\bf
    Q},E)d{\bf Q}}/\int{d{\bf Q}}$ (in units of $\mu_B^2$/eV/f.u.),
where $\chi^{\prime\prime}({\bf Q},E)=(1/3) tr( \chi_{\alpha \beta}^{\prime\prime}({\bf Q},E))$ \cite{clester10,msliu12,mwang13}.

In addition to the TOF INS measurements on MERLIN, we also took data on
the $x=0.096$ compound using the TAIPAN thermal neutron triple-axis spectrometer at the Bragg Institute, Australian Nuclear Science and Technology Organization (ANSTO).
The measurements were carried out on $\sim$29 grams co-aligned single crystals using
the $[H,0,3H] \times [0,-K,0]$ scattering plane \cite{hqluo}.  TAIPAN uses double focusing pyrolytic graphite monochromator and vertical focusing analyzer with
a  pyrolytic graphite filter before the analyzer and
a fixed final neutron energy of $E_f=14.87$ meV.  In $[H,K,3H]$ scattering plane, we performed transverse scans along the $[1, K,3]$ direction for energies up to 25 meV.
The BaFe$_{2-x}$Ni$_x$As$_2$ samples with $x=0.096,0.18$ for TOF and triple-axis experiments are aligned using a \emph{Photonic Sciences} X-ray Laue camera and
co-aligned by using TAIPAN triple-axis spectrometer at ANSTO and ALF crystal alignment facility at ISIS.
The BaFe$_{1.85}$Ni$_{0.15}$As$_2$ samples for TOF experiments are co-aligned using E3 neutron four-circle diffraction
spectrometer at Canadian Neutron Beam Center in Chalker River, Canada.

\begin{figure*}[t]
\includegraphics[width=0.7\textwidth]{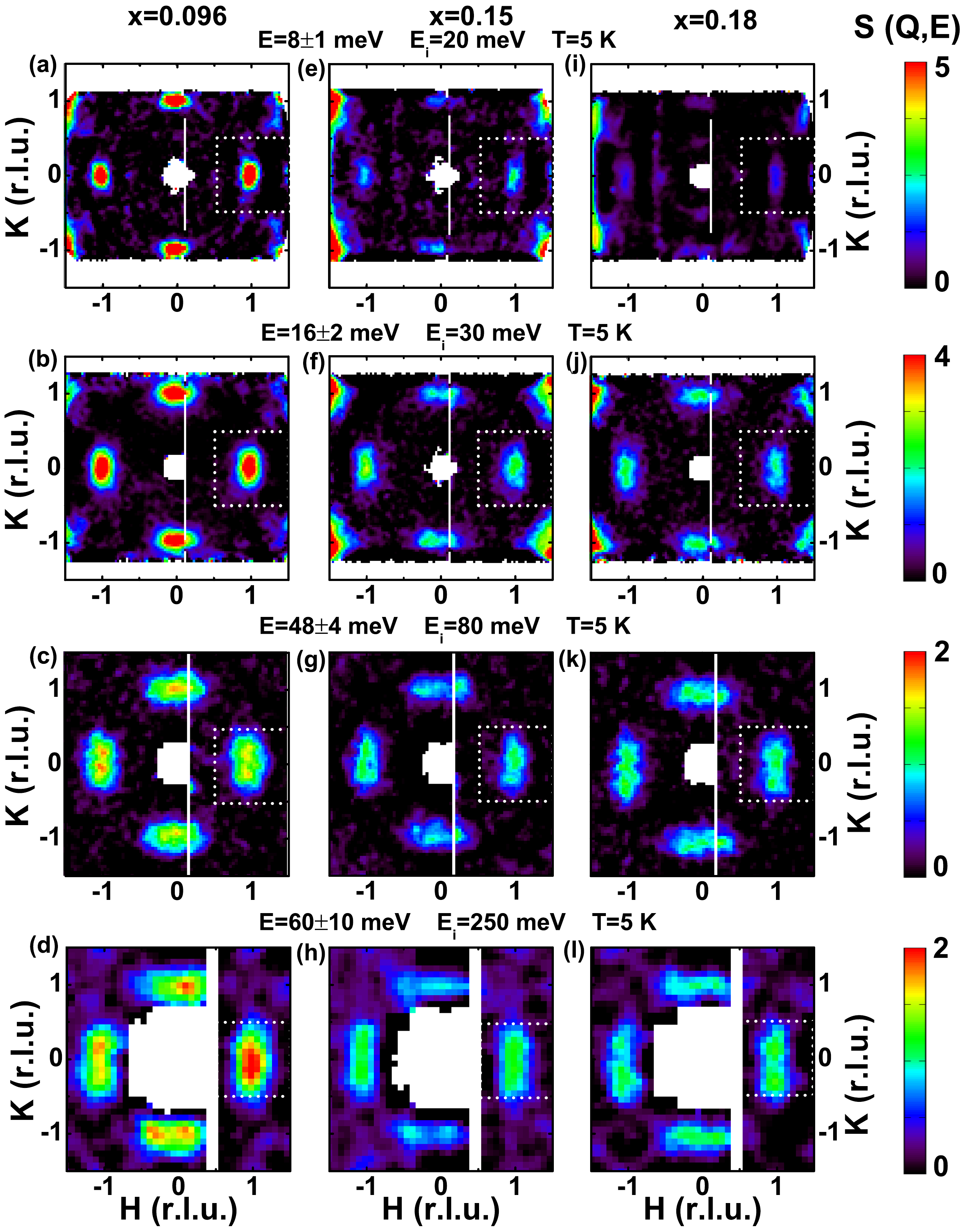}
\caption{(Color online) Comparison of
two dimensional constant-energy slices through the magnetic excitations of BaFe$_{2-x}$Ni$_{x}$As$_{2}$ ($x= 0.096, 0.15$ and 0.18) at
energies of (a,e,i) $E=8\pm 1$, (b,f,j) $16\pm 2$, (c,g,k) $48\pm 4$, and
(d,h,l) $60\pm 10$ meV.  The data in (a,e,i), (b,f,j), (c,g,k), and (d,h,l)
are collected using $E_i= 20, 30, 80, 250$ meV, respectively.
For $E_i\leq 80$ meV, images are obtained after subtracting a radially symmetric $Q$-dependent background
integrated from the diagonal line of the entire zone $-2<H<2$ and $-2<K<2$, which is mainly from
the phonon scattering of the aluminum sample holders.
For $E_i\geq$ 250 meV images are obtained after subtracting the background integrated from
$1.8<H<2.2$ and $-0.2<K<0.2$.  The color bars represent the vanadium normalized
absolute spin excitation intensity in the units of mbarn/sr/meV/f.u. and the dashed boxes indicate AF zone boundaries for
a single FeAs layer.
}
\end{figure*}

\section{Results}

\begin{figure*}[t]
\includegraphics[width=0.7\textwidth]{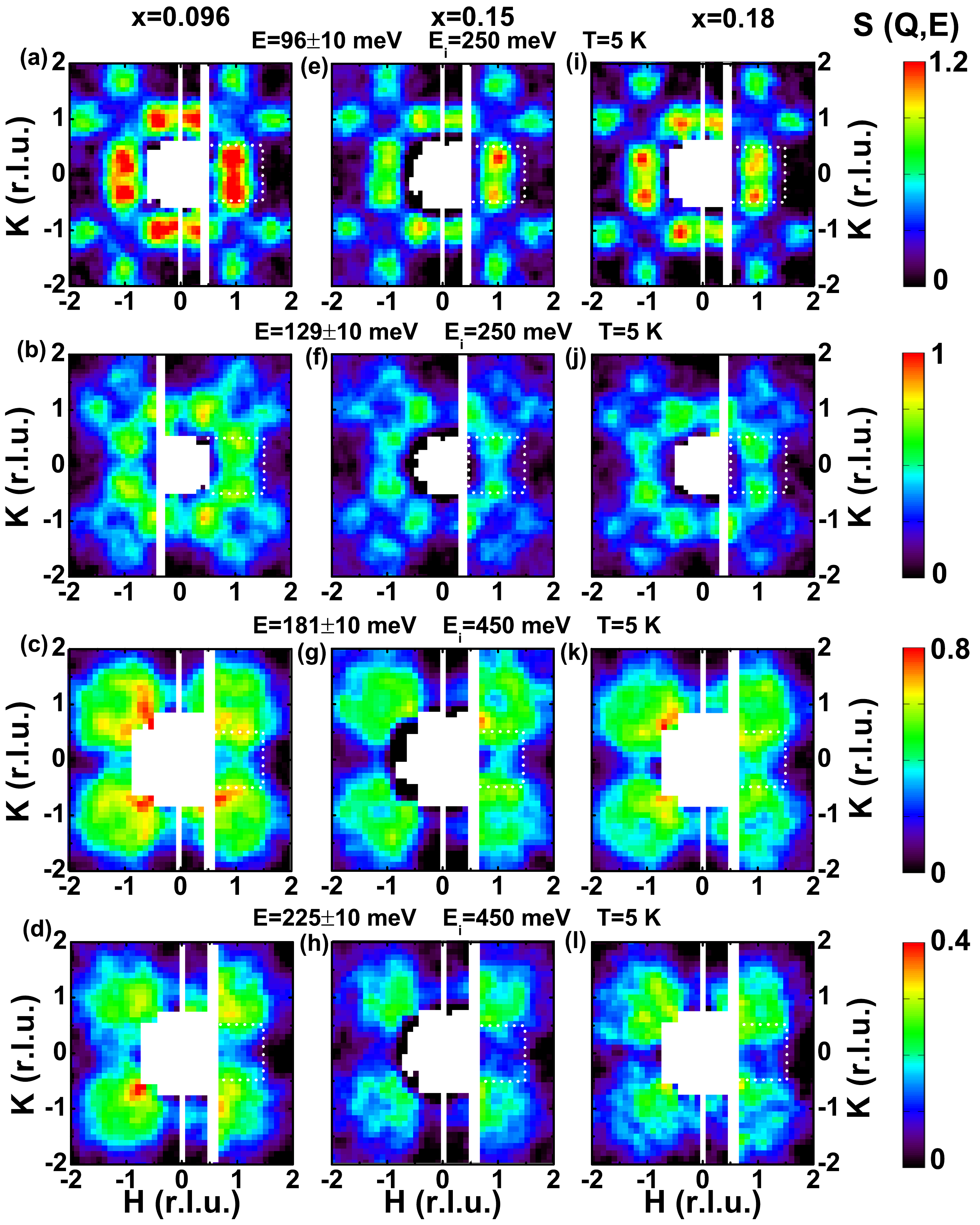}
\caption{
(Color online)
Two dimensional constant-energy slices through the magnetic excitations of BaFe$_{2-x}$Ni$_{x}$As$_{2}$ ($x=$ 0.096, 0.15 and 0.18)
at energies of (a,e,i) $E=96\pm 10$, (b,f,j) $129\pm 10$, (c,g,k) $181\pm 10$,
 and (d,h,l) $225\pm 10$ meV obtained with $E_i= 250$ and 450 meV along the $c$-axis.
}
\end{figure*}
We first describe the evolution of spin excitation disperions
in BaFe$_{2-x}$Ni$_{x}$As$_{2}$.  Figure 1(e) and (f) show the overall
dispersions along the $Q=[1,K]$ and $[H,0]$ directions for $x=$ 0.096, 0.15 and 0.18 compared
with the parent compound $x=0$ (black solid lines).
While the spin excitations at low-energy below $\sim$100 meV become slightly more dispersive upon Ni doping, the high energy spin excitations are not much affected
by electron doping, similar to those
in the heavily electron overdoped
BaFe$_{1.7}$Ni$_{0.3}$As$_{2}$.  These results suggest that
the effective magnetic exchange couplings $J$ are not much affected by electron-doping in
BaFe$_{2-x}$Ni$_{x}$As$_{2}$ for $x\leq 0.3$ \cite{mwang13}.

To further study the effect of electron-doping to the low energy ($E< 60$ meV) spin excitations,
we fit the wave vector dependence of the spin excitations along the $[1,K]$ and $[H,0]$ directions by Gaussian on a linear background to
estimate their full-width-at-half-maximum (FWHM) in
BaFe$_{2-x}$Ni$_x$As$_2$ with $x=0.096, 0.15, 0.18$.
For comparison, we also probe the low-energy spin excitations on BaFe$_{2-x}$Ni$_x$As$_2$ with $x=0.096$ using TAIPAN triple-axis spectrometer along the
$Q=[1, K, 3]$ direction.
The solid black lines in Figs. 2(a) and 2(b) are spin wave dispersions of BaFe$_2$As$_2$ estimated using the previous obtained in-plane
effective magnetic exchange couplings \cite{lharriger} and appropriate spin anisotropy gap values \cite{jtpark12}.
The shaped area in Figs. 2(a) and 2(b) show the FWHM of spin excitations
for BaFe$_{2-x}$Ni$_x$As$_2$ with $x=0.096, 0.15, 0.18$ along the $[1,K]$ and $[H,0]$ directions, respectively.
Figure 2(c) shows energy dependence of the spin excitation widths along the $[1,K]$ direction.
Within the probed energy range ($3\leq E\leq 60$ meV), the widths of spin excitations
increase monotonically with increasing energy and the electron-doping level $x$ for these three samples.
For BaFe$_{1.904}$Ni$_{0.096}$As$_2$, the spin excitation widths determined from the
triple-axis experiments are slightly smaller than that from the TOF measurements due to the differences
in the instrumental resolutions in these two techniques.  Figure 2(d) shows the energy dependence of the spin excitation widths along the $[H,0]$ direction, which
are almost independent of electron-doping above 30 meV.  Thus the low-energy spin excitations are transversely elongated upon doping
and become broader than spin waves in the undoped compound.

\begin{figure*}[t]
\includegraphics[width=0.6\textwidth]{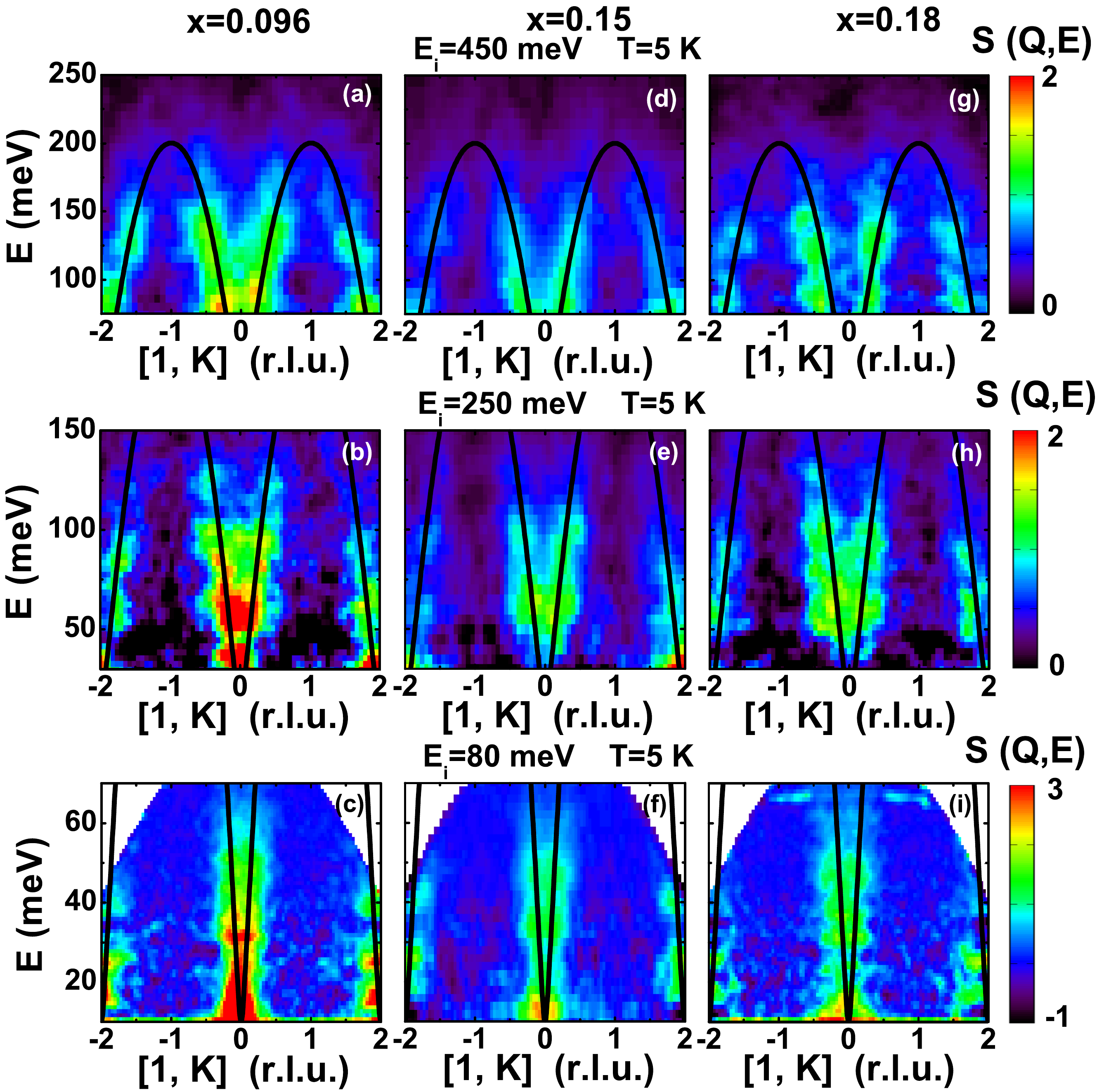}
\caption{
(Color online)
Energy dependence of the two-dimensional slices along the
$Q=[1, K]$ direction
with $E_i= 80$, 250 and 450 meV for panels (c,f,i), (b,e,h), and (a,d,g),
respectively.
The solid lines are dispersions of
spin waves in BaFe$_{2}$As$_{2}$ \cite{lharriger}.
}
\end{figure*}

To directly compare the evolution of spin excitations as a function of increasing electron-doping $x$, we show in Figs. 3 and 4 TOF INS measurements for
$x=0.096,0.15,0.18$ obtained on MERLIN using identical setup.  The scattering intensity is normalized to absolute units of mbarn/sr/meV/f.u. using a
vanadium standard and the dashed boxes mark the AF Brillouin zone for the magnetic unit cell with single Fe$^{2+}$.  For energies below 70 meV
[$E=8\pm 1$ meV, Figs. 3(a), 3(e), 3(i); $16\pm 2$ meV, Figs. 3(b), 3(f), 3(j); $48\pm4$ meV, Figs. 3(c), 3(g), 3(k); $60\pm10$ meV,
Figs. 3(d), 3(h), 3(l) for $x=0.096,0.15$, and 0.18, respectively], spin excitations are transversely elongated ellipses
centered around the in-plane AF ordering wave vectors ${\bf Q}_{AF}=(\pm1,0)$ and $(0,\pm 1)$ due to the two twinned domains.
The excitations become more transversely elongated and decrease in intensity
with increasing $x$.  On increasing energies to $E=96\pm10$ meV [Figs. 4(a), 4(e), and 4(i)
for $x=0.096,0.15$, and 0.18, respectively], spin excitations starts to split transversely away from the AF ordering wave vectors and become less doping dependent.
For energies $E=129\pm10$ meV [Figs. 4(b), 4(f), and 4(j)], $E=181\pm10$ meV [Figs. 4(c), 4(g), and 4(k)], and $E=225\pm10$ meV [Figs. 4(d), 4(h), and 4(l)],
spin excitations become rather similar, and are almost electron-doping independent.

\begin{figure*}[t]
\includegraphics[width=0.6\textwidth]{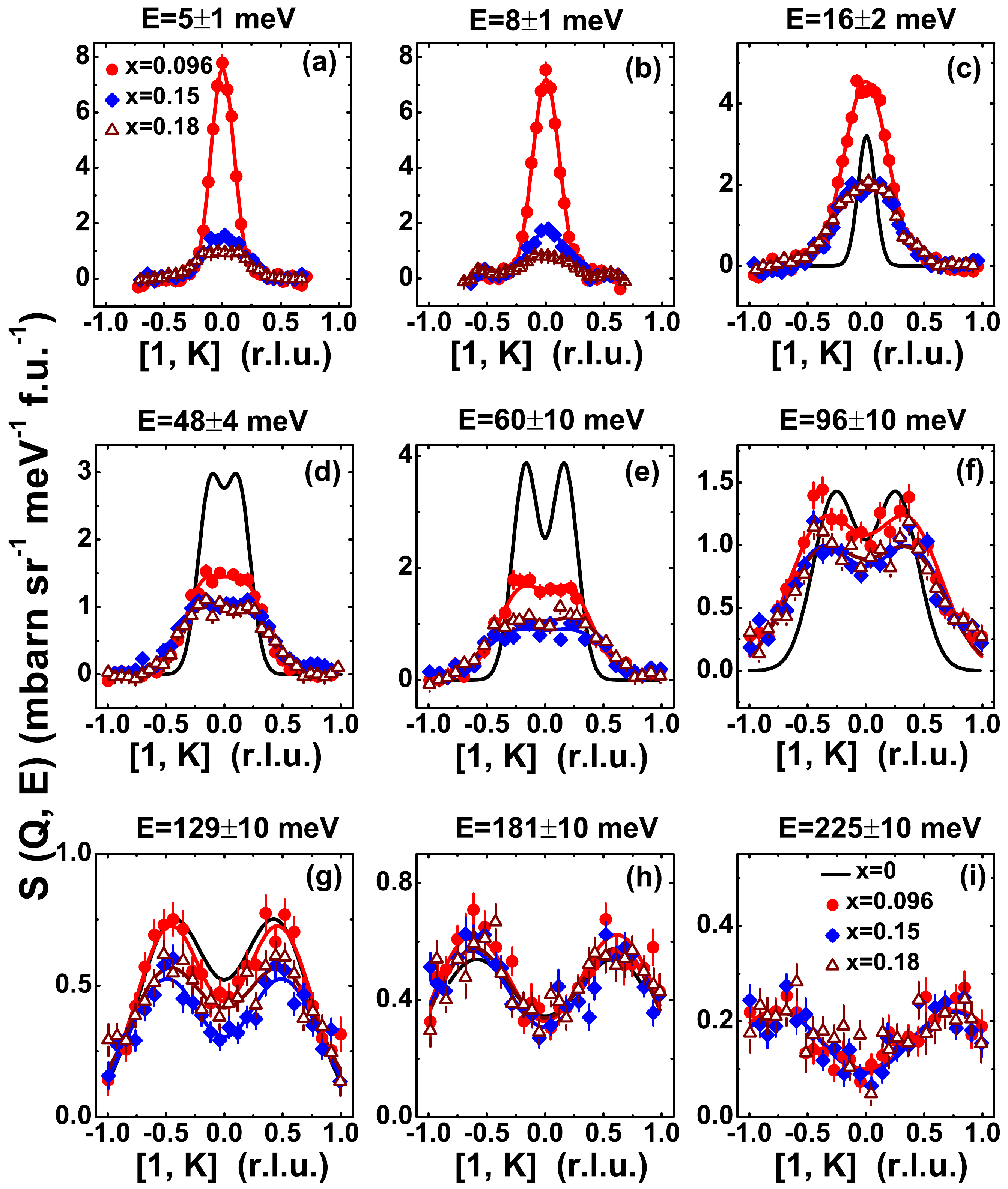}
\caption{
(Color online)
Constant-energy cuts in the spin excitations of BaFe$_{2-x}$Ni$_{x}$As$_{2}$ along the $[1,K]$ direction
at different energies corresponding to those in Fig. 3 and Fig. 4, where the wave vector integration ranges are $0.9<H<1.1$ for the $K$ cuts
and $-0.1<K<0.1$ for the $H$ cuts.
The solid lines are Gaussian fitting results for each doping and
spin waves in the parent compound BaFe$_{2}$As$_{2}$ \cite{lharriger}.
}
\end{figure*}

Figure 5 compares the background subtracted scattering for the
$E_i = 450, 250,$ and 80 meV data projected in the wave vector ($Q = [1, K]$)
and energy space for  BaFe$_{2-x}$Ni$_x$As$_2$ with $x=0.096, 0.15$, and 0.18.
These incident beam energies were chosen to probe  spin excitations at different energies.
Figures 5(a), 5(d), and 5(g) show the $E_i=450$ meV data for the $x=0.096,0.15$, and 0.18 samples, respectively.
Similar data with $E_i=250$ and 80 meV are shown in Figs. 5(b), 5(e), 5(h), and
5(c), 5(f), 5(j), where the solid lines are spin wave dispersions for BaFe$_2$As$_2$ \cite{lharriger}.
While magnetic scattering clearly decreases with increasing doping at energies below 60 meV, they are virtually unchanged
for energies above 100 meV, consistent with results in Figs. 3 and 4.  To quantitatively determine the evolution of spin
excitations for BaFe$_{2-x}$Ni$_{x}$As$_{2}$ with $x=0,0.096,0.15$, and 0.18, we show in Figs. 6 and 7 constant-energy cuts at different energies along the
$[1,K]$ and $[H,0]$ directions, respectively.  At $E=5\pm1$ [Fig. 6(a)] and $8\pm 1$ meV [Fig. 6(b)], the commensurate spin excitations at $x=0.096$ become
weaker and transversely incommensurate on moving to $x=0.15,0.18$.  For energies of $E=16\pm 2$ [Fig. 6(c)], $48\pm4$ [Fig. 6(d)], and $60\pm10$ [Fig. 6(e)] meV,
the electron-doping induced spin excitation intensity reduction becomes smaller.  Finally, there are no significant difference
between spin excitations of the parent compound and $x=0.096,0.15,0.18$ at $E=96\pm10$ [Fig. 6(f)], $129\pm10$ [Fig. 6(g)], $181\pm10$ [Fig. 6(h)], and
$225\pm10$ meV [Fig. 6(i)].  Figures 7(a)-7(d) show the comparison of $[H,0]$ scans for the $x=0.096,0.15$, and 0.18 samples at $E=16\pm2$, $48\pm4$, $96\pm10$, and
$129\pm10$ meV.  While the electron-doping evolution of the spin excitation intensity is consistent with cuts along the $[1,K]$ direction, they
are commensurate at all energies probed.

\begin{figure}[t]
\includegraphics[scale=.3]{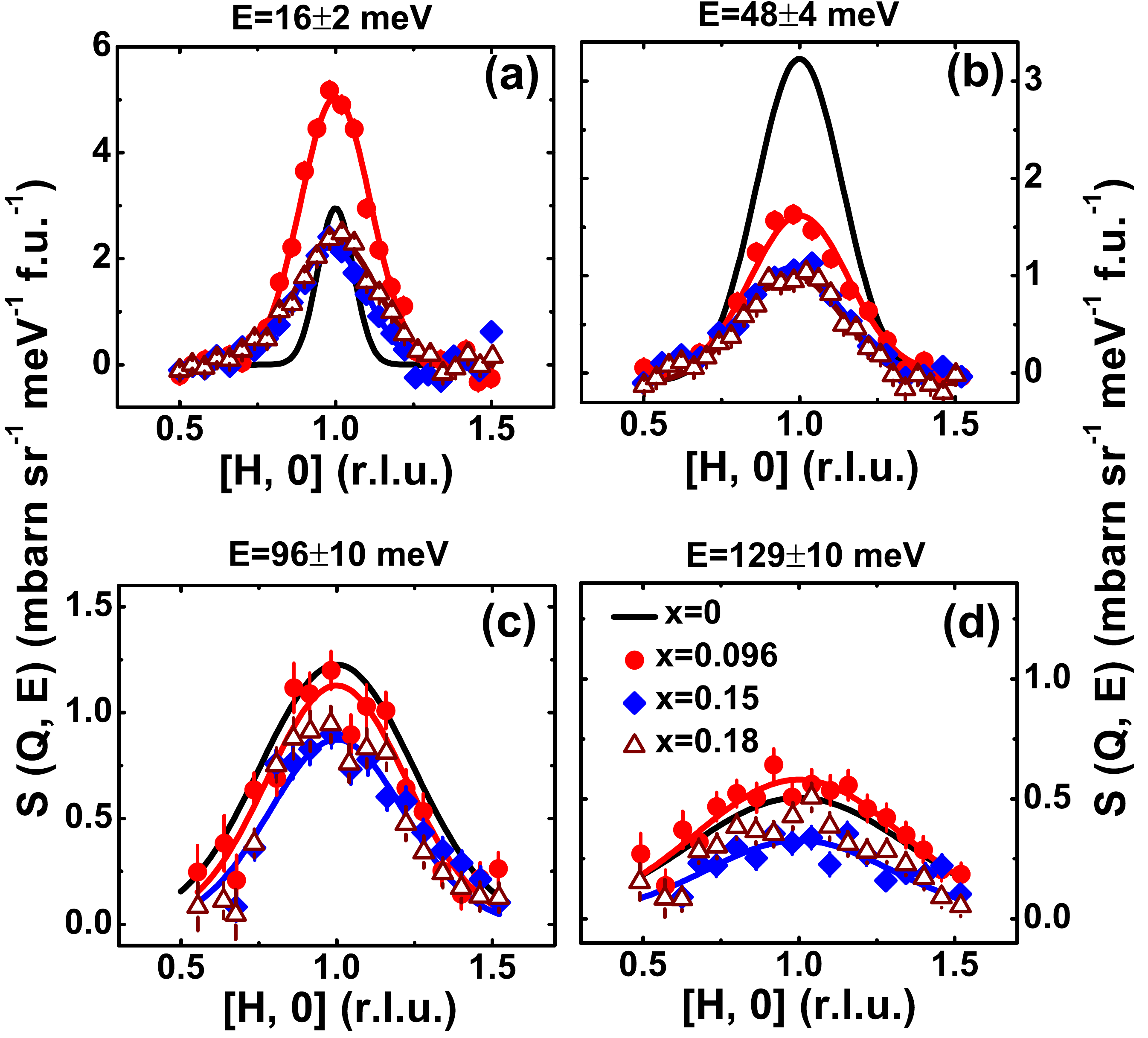}
\caption{(Color online)
Constant-energy cuts of the spin excitations in
BaFe$_{2-x}$Ni$_{x}$As$_{2}$ along the $[H,0]$ direction at (a) $E=16\pm 2$ meV, (b) $48\pm4$ meV,
(c) $96\pm10$ meV, and $129\pm10$ meV.  The solid lines are spin wave cuts in parent compound, and the reduced
intensity compared with the doped material at $E=16\pm2$ meV is due to the presence of a large spin anisotropy
gap \cite{jtpark12}.  The spin excitations are commensurate at all energies probed.
}
\end{figure}

To illustrate further the electron-doping evolution of the spin excitations in the overdoped regime, we
compare constant-$Q$ cuts in spin excitations
of BaFe$_{2-x}$Ni$_x$As$_2$ with $x=0,0.096,0.15$, and 0.18 in Fig. 8 \cite{msliu12}.
The arrows in the inset of Fig. 8(a) show the directions of the constant-$Q$ cuts.
At wave vectors near the Brillouin zone center at $Q=(1,0.05)$ and $(1,0.2)$,
electron-doping clearly suppresses the low-energy spin excitations.  On increasing the wave vector
to $Q=(1,0.35)$ and $(1,0.5)$, there are much less difference in spin excitations of undoped and doped
materials.  Spin excitations form a broad peak near 100 meV
in electron-overdoped samples similar to spin waves in parent compound.

Having established the electron-doping evolution of the overall spin excitations spectra, we now describe the effect of superconductivity on the
low-energy spin excitations.
From previous work, we know that a neutron spin resonance appears in the superconducting state of iron pnictides \cite{mdlumsden,sxchi,sli09,dkpratt09,adChristianson,dsinosov09,mywang10,clester10,jtpark10,hfli10,dsinosov11,mwang11}.
Careful temperature dependent study of the resonance in the superconducting
BaFe$_{1.85}$Co$_{0.15}$As$_2$ iron pnictide suggests that the mode energy decreases on warming to $T_c$ and is coupled with the decreasing
superconducting gap energy \cite{dsinosov09}.
This is different from the resonance in superconducting iron
chalcogenide Fe$_{1+\delta}$Te$_{1-x}$Se$_x$, where the mode energy is
weakly temperature dependent \cite{mook,qiu,lumsden2,shlee,dnargyriou,slli10,zjxu,leland12}.
Very recently, a sharp neutron spin resonance has been identified in
superconducting NaFe$_{0.935}$Co$_{0.045}$As ($T_c=18$ K) iron pnictide \cite{clzhang13}.  Here, the resonance energy is again found to be weakly temperature dependent similar to
the mode in Fe$_{1+\delta}$Te$_{1-x}$Se$_x$ \cite{clzhang13}.  In order to probe the detailed temperature dependence of the resonance in
BaFe$_{1.904}$Ni$_{0.096}$As$_2$, we carried out TOF INS measurements on MERLIN with $E_i=30$ meV at many temperatures below and above $T_c$.
Following previous practice \cite{mdlumsden,sxchi,sli09,dkpratt09,adChristianson,dsinosov09,mywang10,clester10,jtpark10,hfli10,dsinosov11,mwang11}, we used the
$T=25$ K data as background and assumed that the net intensity gain near the AF ordering wave vector at lower temperatures is the resonance.  Since
an incident beam energy of $E_i=30$ meV corresponds to $L\approx 1$ r.l.u. near the resonance energy of $E_r\approx 7$ meV, we can simultaneously probe
the wave vector and energy dependence of the mode below $T_c$.   Figures 9(a)-9(f) show the wave vector dependence of the
temperature differences (the low-temperature data subtracts the data at 25 K) in
spin excitations, $S({\bf Q},E,T)-S({\bf Q},E,T=25\ {\rm K})$ with $E_r=7\pm1$ meV,
 at $T=5,11,14,16,18,20$ K, respectively.  At $T=5$ K, the superconductivity-induced resonance forms a transversely elongated ellipse in the
 $[H,K]$ plane centered at ${\bf Q}_{AF}=(1,0)$ [Fig. 9(a)].  On warming to $T=11$ K [Fig. 9(b)], 14 K [Fig. 9(c)],
 and 16 K [Fig. 9(d)], the resonance  becomes weaker and broader along both the
 $[H,0]$ and $[1,K]$ directions.  The resonance becomes almost indistinguishable from the background at $T=$ 20 K [Fig. 9(f)].

\begin{figure}[t]
\includegraphics[scale=.3]{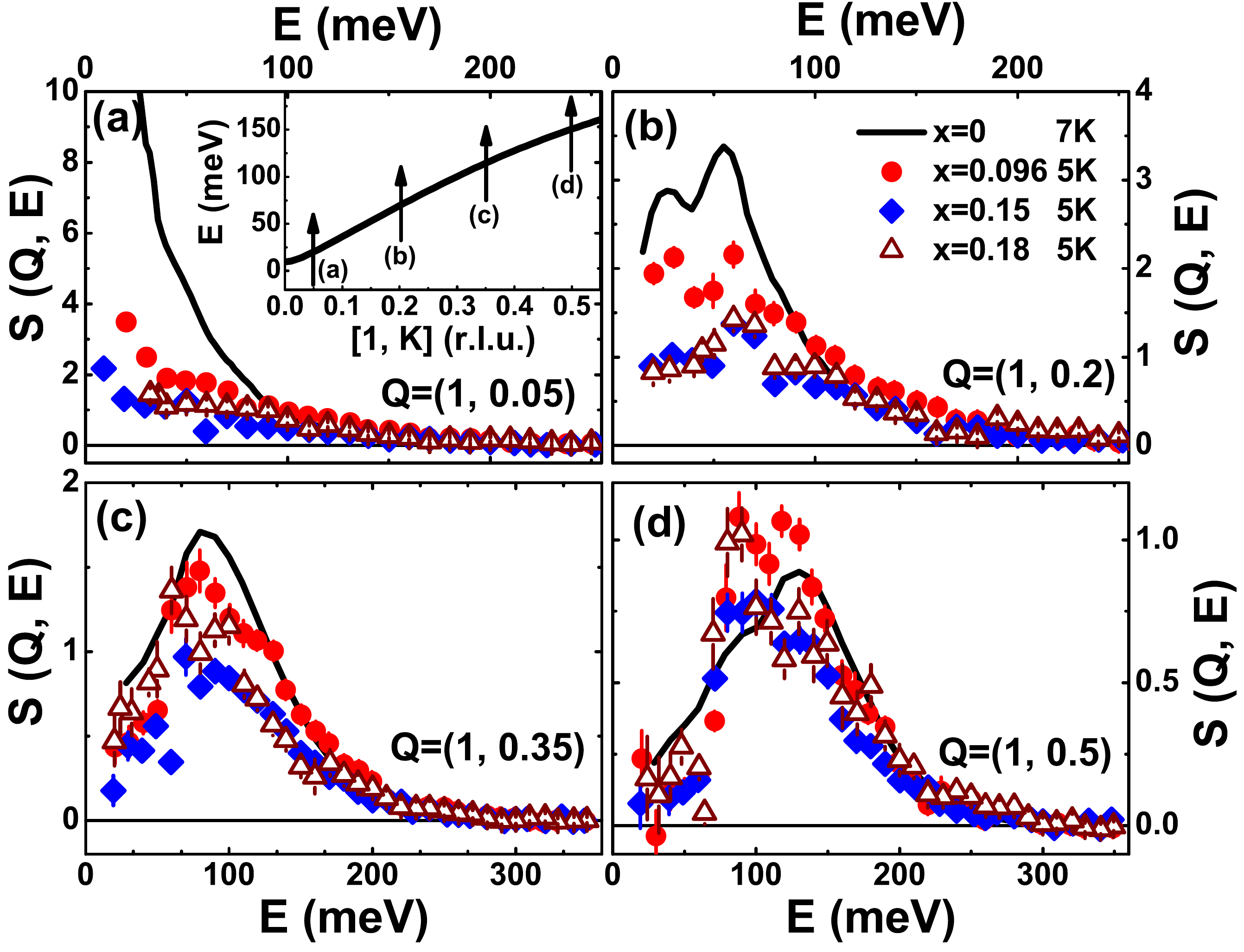}
\caption{(Color online)
Constant-$Q$ cuts in the spin excitations of BaFe$_{2-x}$Ni$_{x}$As$_{2}$ at wave vectors
(a) $Q= (1, 0.05)$, (b) (1, 0.2), (c) (1, 0.35), and (d) $(1, 0.5)$
as marked by the vertical arrows in the inset of (a).
The solid lines are identical cuts from spin waves in BaFe$_2$As$_2$ \cite{lharriger}.
}
\end{figure}

Figures 9(g)-9(l) show the net magnetic scattering above the $T=25$ K
background projected onto the $[1,K]$ and energy space at different temperatures.
At $T=5$ K, we see a clear neutron spin resonance centered at $E_r=7\pm1$ meV and ${\bf Q}_{AF}=(1,0)$ [Fig. 9(g)].  Although the intensity
of the resonance becomes progressively weaker on warming up to temperatures $T=11$ K [Fig. 9(h)], 14 K [Fig. 9(i)], and 16 K [Fig. 9(j)],
its peak position in energy appears to be fixed at $E\approx 7$ meV.  On further warming to $T=18$ K [Fig. 9(k)],
one can still see a weak resonance near $E_r\approx 7$ meV.  It becomes impossible to decern any magnetic signal
at $T=20$ K above the $T=25$ K background scattering [Fig. 9(l)].

To quantitatively determine the temperature evolution of the
resonance, we cut the images in Figs. 9(g)-9(l) along the energy direction by integrating wave vectors $0.8<H<1.2$ and $-0.2<K<0.2$ r.l.u. around
 ${\bf Q}_{AF}=(1,0)$.    Figure 10(a) shows the outcome at temperatures
 in Fig. 9(g)-9(l) and additional data taken at $T=19$ K and 22 K.
At all temperatures below $T_c=20$ K, we see a well-defined resonance showing as positive scattering above background
near $E= 7$ meV.  There are no statistical differences in magnetic scattering for temperatures
between $T=20,22$ K and 25 K.  Figure 10(b) shows the wave vector cuts along the $[1,K]$
direction with energy-integration of
$E=7\pm 1$ meV and ${\bf Q}$-integration from $0.8<H<1.2$
 at different temperatures.  There are well-defined peaks centered at the commensurate AF ordering wave vector for all probed temperatures.
The solid lines are Gaussian fits to the data, which give peak intensity and FWHM of the spin excitations.
Figure 10(c) shows similar wave vector cuts along the $[H,0]$ direction with Gaussian fits.  The superconductivity-induced effects on wave vector dependence of the
resonance along the $[1,K]$ and $[H,0]$ directions are shown in Figs. 10(d) and 10(e), respectively.  The data are peaked around the
${\bf Q}_{AF}=(1,0)$ wave vector and the solid lines are Gaussian fits on zero backgrounds.

\begin{figure*}[t]
\includegraphics[width=0.5\textwidth]{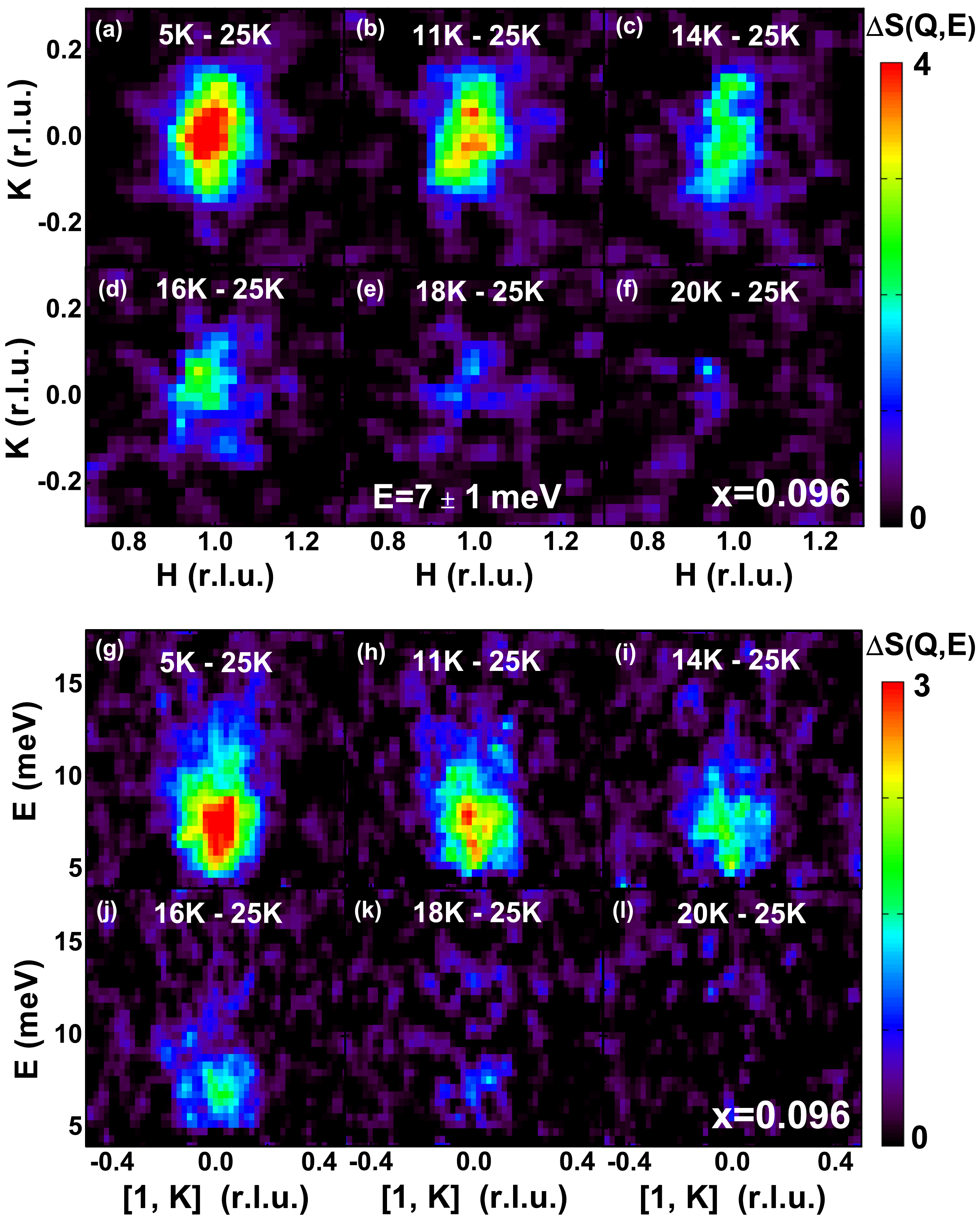}
\caption{
(Color online)
(a-f) The wave vector dependence of the resonance in
BaFe$_{1.904}$Ni$_{0.096}$As$_2$ at $T=5$ K, 11 K, 14 K, 16 K, 18 K and 20 K after subtracting the normal state data at 25 K.
(g-l) Energy dependence of the two-dimensional slices along the $Q=[1, K]$ direction
for the resonance at different temperatures.  The mode essentially disappears around 20 K, but its peak positions are weakly
temperature dependent.
}
\end{figure*}

\begin{figure}[t]
\includegraphics[width=0.4\textwidth]{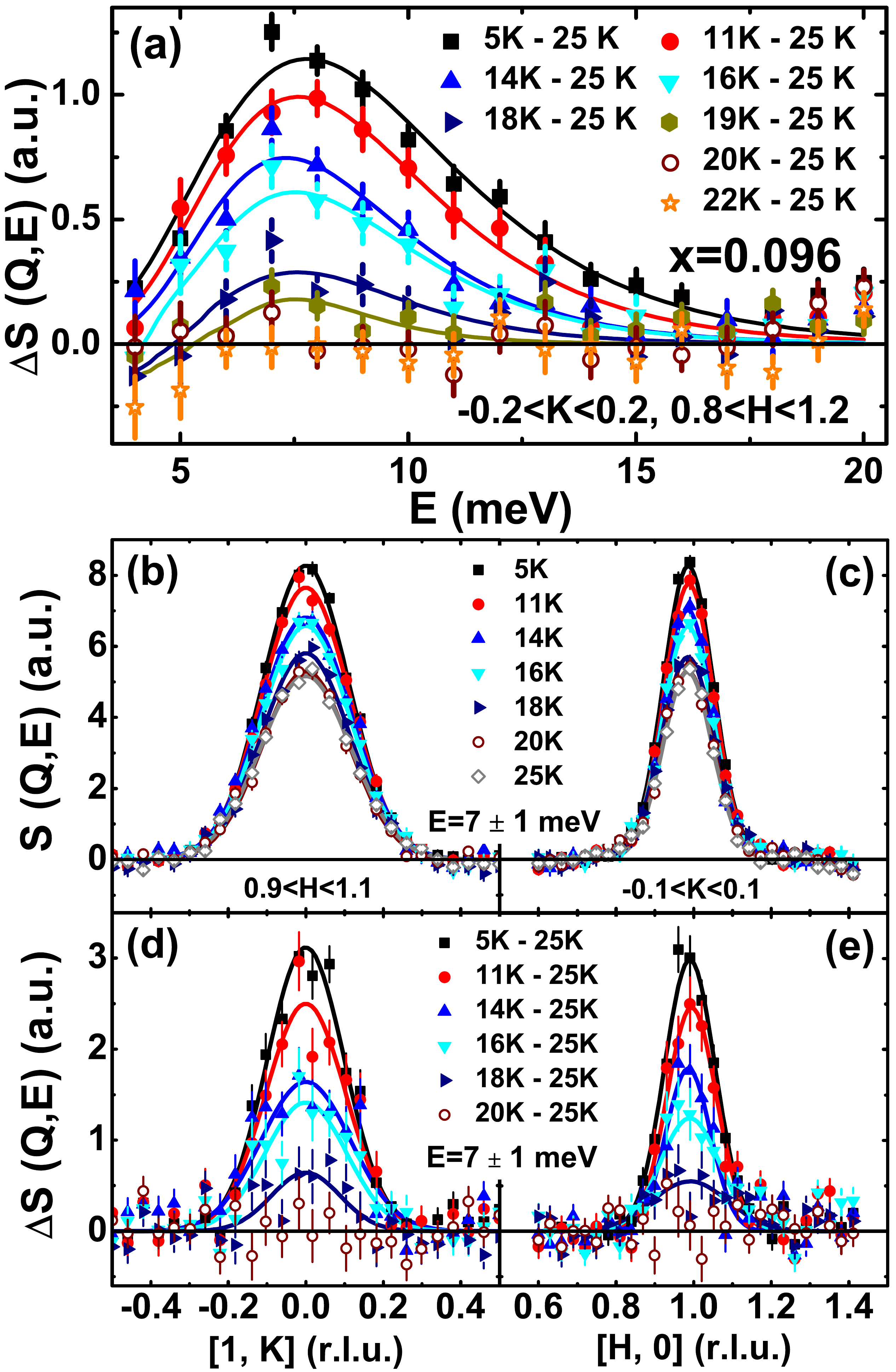}
\caption{(Color online)
Temperature dependence of the resonance in
BaFe$_{1.904}$Ni$_{0.096}$As$_2$
 measured with $E_i=20$ meV.
 (a) The difference of constant-$Q$ cuts between the low-energy spin excitations at $T< 25$ K and normal state at $T=25$ K around AF ordering
  wave vector with integration range $0.8<H<1.2$ and $-0.2<K<0.2$.
  (b,c) Constant-energy cuts along the $Q=[1,K]$ and $[H,0]$ directions
  at different temperatures and $E=7\pm 1$ meV. (d) and (e) Wave vector dependence of the resonance at $E=7\pm 1$ meV.
}
\end{figure}

\begin{figure}[t]
\includegraphics[width=0.35\textwidth]{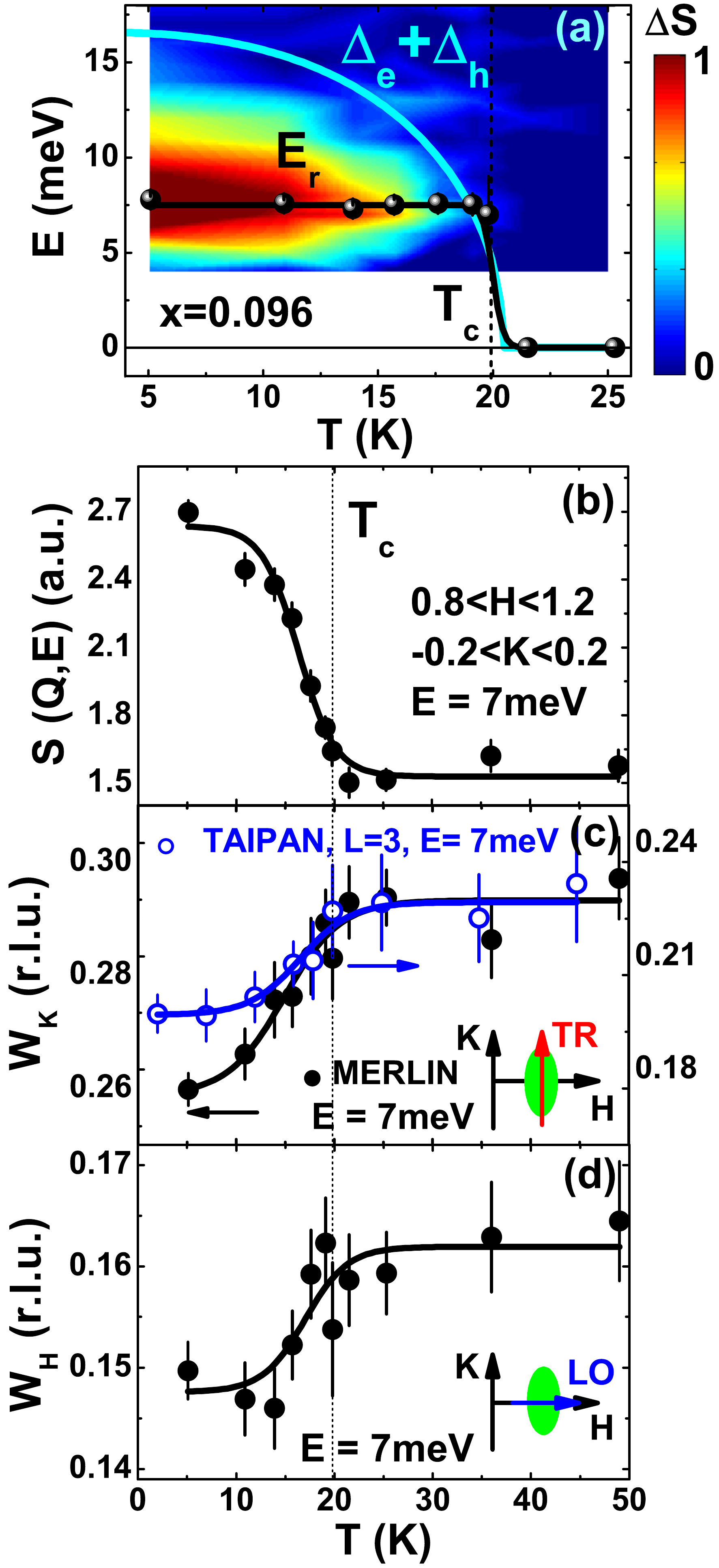}
\caption{(Color online)
(a) Temperature dependence of the resonance
energy $E_r$ and the sum of superconducting gaps $\Delta_e+\Delta_h$
for BaFe$_{1.904}$Ni$_{0.096}$As$_2$
 \cite{zswang}. The color bars represent the intensity gain $\Delta S(Q,E)$ around AF ordering
 wave vector in the superconducting state. (b) Temperature dependence of the integrated intensity of the
 spin excitations at $E=$ 7 meV with integration range $0.8<H<1.2$ and $-0.2<K<0.2$.
 (c,d) Temperature dependence of FWHM (peak width) along the
 $[1,K]$ ($W_K$) and $[H,0]$ ($W_H$) directions at $E=7$  meV.
 The blue open circles are the similar results from the triple-axis experiments at fixed $L=3$.
}
\end{figure}

Using parameters obtained from fits to the spin excitations spectra in Fig. 10, we can determine the temperature dependence of the resonance energy, intensity, and
${\bf Q}$-widths along the $[1,K]$ and $[H,0]$ directions.
These results can be compared with temperature dependence of the superconducting gaps determined from other methods \cite{terashima,tanatar10,zswang}.
From angle resolved photoemission spectroscopy experiments \cite{terashima}, it is well known that
the electron-doped BaFe$_{2-x}T_x$As$_2$ iron pnictides have the large isotropic superconducting gaps $\Delta_h$ located on the
hole Fermi surface near the zone center position $\Gamma$
and the small gap $\Delta_e$ on one of the electron Fermi surfaces near $M$ point.
The temperature dependence of the superconducting gaps decrease
with increasing temperature and vanish at $T_c$.
The pink solid line in Figure 11(a) shows temperature dependence of the sum of the electron and hole Fermi surface
superconducting gaps $\Delta_e+\Delta_h$ obtained from point-contact andreev reflection measurements on
BaFe$_{1.9}$Ni$_{0.1}$As$_2$ \cite{zswang}.  By comparing the temperature dependence of the resonance in
the color contour plot and the solid points with the pink solid line, we see that the energy position of the resonance
is weakly temperature dependent and does not follow the temperature dependence of
the sum of the electron and hole pocket superconducting gaps.  This is similar to
the temperature dependence of the resonance in superconducting iron
chalcogenide Fe$_{1+\delta}$Te$_{1-x}$Se$_x$ \cite{qiu,leland12} and
NaFe$_{0.935}$Co$_{0.045}$As iron pnictide \cite{clzhang13}.

\begin{figure}[t]
\includegraphics[scale=.29]{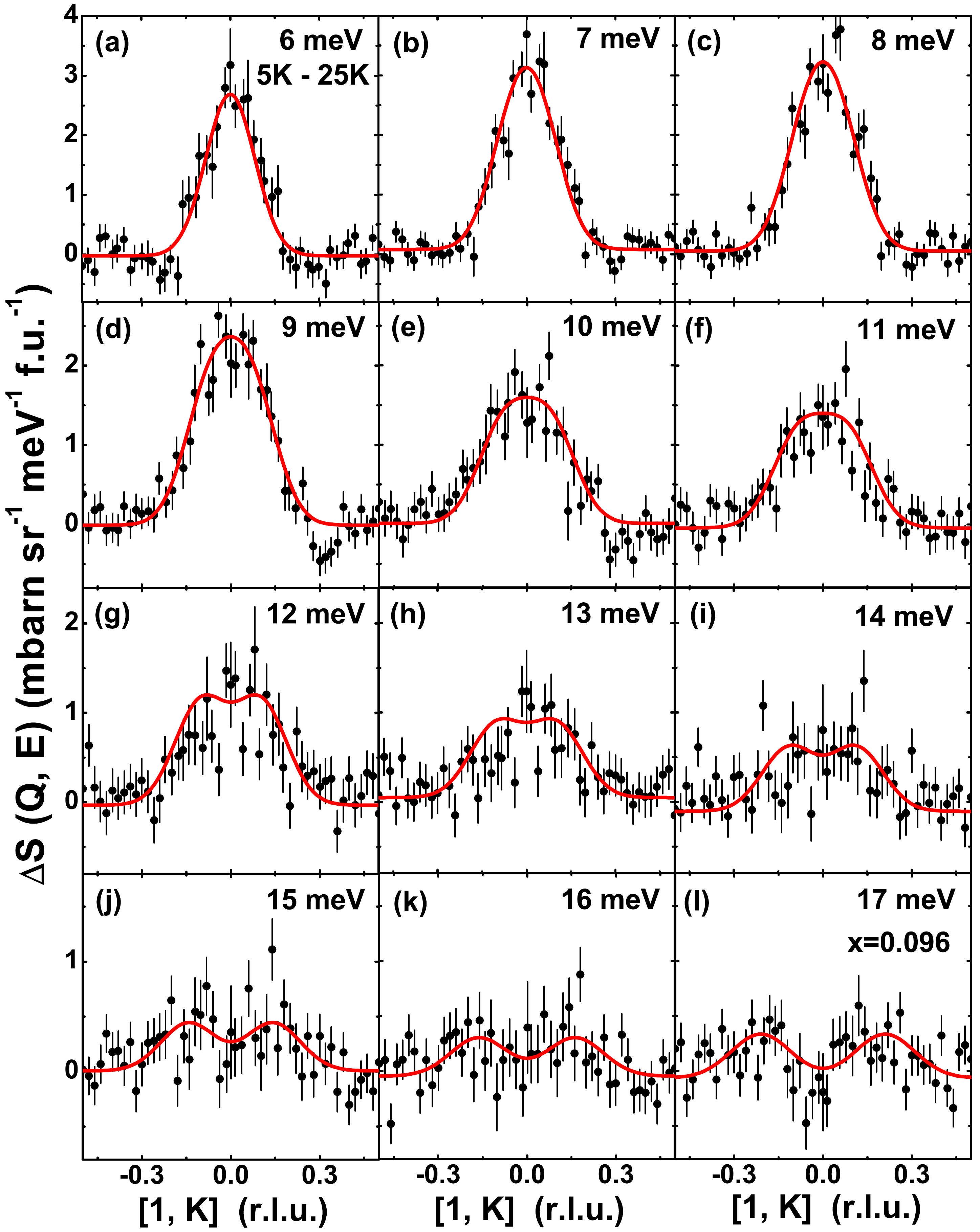}
\caption{(Color online)
Wave vector dependence of the resonance as a function of increasing energy for BaFe$_{1.904}$Ni$_{0.096}$As$_2$.
Using the color plot in Fig. 9(g), we
cut the data long the $[1,K]$ direction for energies of (a) $E=6\pm1$, (b) $7\pm1$,
(c) $8\pm1$, (d) $9\pm1$, (e) $10\pm1$, (f) $11\pm1$, (g) $12\pm1$, (h) $13\pm1$,
(i) $14\pm1$, (j) $15\pm1$, (k) $16\pm1$, (l) $17\pm1$ meV.  The solid lines are fitting results by two symmetric Gaussian functions on flat backgrounds.
}
\end{figure}

Figure 11(b) shows the temperature dependence of the wave vector integrated ($0.8<H<1.2$ and $-0.2<K<0.2$) magnetic spectral
weight.  Consistent with earlier measurements \cite{hqluo12}, the spectral weight of the resonance increases below
$T_c$ like an order paramter of superconductivity.  The filled solid circles in Figs. 11(c) and 11(d) plot temperature dependence of FWHM of the resonance along the transverse
and longitudinal directions, respectively.
The open circles in Fig. 11(c) show temperature dependence of the FWHM of the resonance
obtained on TAIPAN triple-axis spectrometer.
We see that
the widths of the resonance display a clear superconductivity-induced narrowing below $T_c$ along both the transverse and longitudinal directions.
In the case of hole-doped iron pnictide superconductor Ba$_{0.67}$K$_{0.33}$Fe$_2$As$_2$ ($T_c=38$ K),
the resonance at $E_r=15$ meV has a longitudinally elongated line shape around the AF ordering wave vector
 ${\bf Q}_{AF}=(1,0)$ in the superconducting state \cite{clzhang11}.  Upon warming across $T_c$, the resonance become isotropic circle in
 reciprocal space \cite{mwang13}.  Such behavior is different from the resonance in electron-doped
BaFe$_{1.904}$Ni$_{0.096}$As$_2$ superconductor, where the transversely elongated
spin excitations become slightly narrower below $T_c$.  In a recent INS experiment on superconducting
BaFe$_{1.926}$Ni$_{0.074}$As$_2$ ($T_c=17$ K) \cite{mgkim13}, the resonance at $E_r=6$ meV was found to have spin wave like dispersion along the
transverse direction.  In our TOF INS measurements for BaFe$_{1.904}$Ni$_{0.096}$As$_2$, this would correspond to
a dispersive
resonance along the transverse $[1,K]$ direction in Fig. 9(g).
To see if we can detect the possible dispersion of the resonance, we cut the temperature difference plot in Fig. 9(g) along the $[1,K]$ direction in 1 meV interval.
The outcome in Fig. 12 shows that the resonance indeed disperses outward for energies above $\sim$10 meV along the transverse direction.
This result, combined with earlier observation of incommensurate resonance in electron overdoped
BaFe$_{1.85}$Ni$_{0.15}$As$_2$ \cite{hqluo12}, indicate that the transversely dispersive resonance mode is prevalent in both the electron
underdoped \cite{mgkim13} and overdoped BaFe$_{2-x}$Ni$_{x}$As$_{2}$.  At present, it is unclear how to understand
the wave vector
narrowing of the resonance below $T_c$ [Figs. 11(c) and 11(d)] at $E=7$ meV and
the dispersion of the mode at higher energies from Fermi surface nesting point of view \cite{hqluo12,maier08,maier09,Korshunov}.

\begin{figure}[t]
\includegraphics[scale=.3]{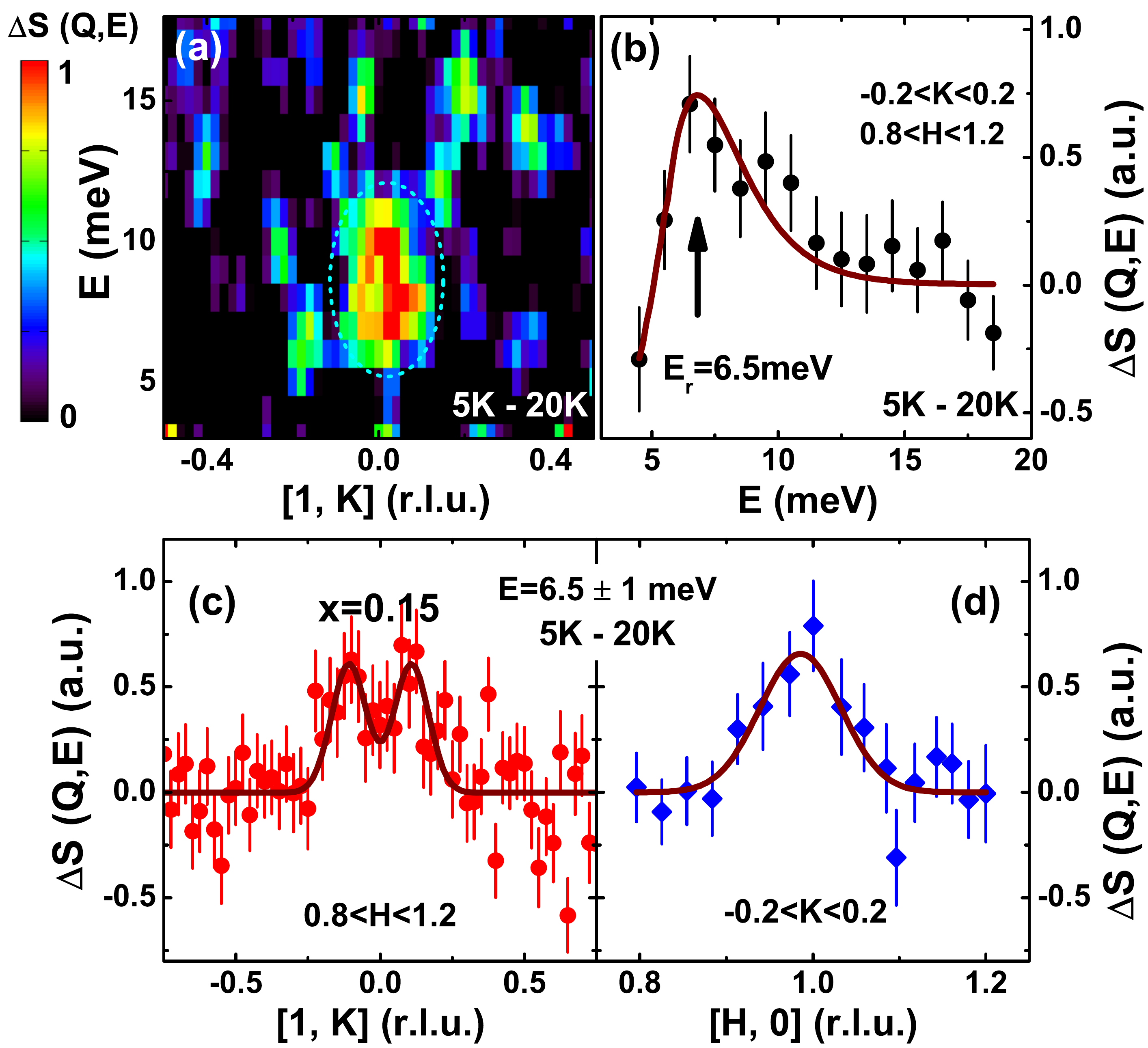}
\caption{(Color online)
The temperature difference plot showing the presence of a resonance near $E_r=6.5$ meV projected
onto the energy-$[1,K]$ plane for BaFe$_{1.85}$Ni$_{0.15}$As$_2$. The data was taken using
 $E_i=25$ meV with incident beam along the $c$-axis.
 (a) The two-dimensional image of the spin excitations
between $T=5$ K and 20 K. (b) Intensity gain of the resonance obtained by integrating
$0.8<H<1.2$ and $-0.2<K<0.2$.  The mode occurs at $E_r=6.5$ meV at 5 K.
(c)
Wave vector dependence of the resonance showing incommensurability along the
$[1, K]$ direction. (d)
The resonance is commensurate along the $[H,0]$ direction.
}
\end{figure}

Having described the temperature, wave vector and energy dependence of the low-energy spin excitations in BaFe$_{1.904}$Ni$_{0.096}$As$_2$, we now
discuss similar TOF INS measurements
for BaFe$_{1.85}$Ni$_{0.15}$As$_2$.  In previous triple-axis and TOF INS measurements on BaFe$_{1.85}$Ni$_{0.15}$As$_2$ \cite{hqluo12},
an incommensurate neutron spin resonance has been identified.  Figure 13(a) shows the temperature difference of spin excitations
between 5 K  and 20 K projected onto the energy and $[1,K]$ plane.
By integrating wave vectors from $0.8<H<1.2$ and $-0.2<K<0.2$, we plot
the energy dependence of the resonance in
Figure 13(b).  The mode energy is now at $E_r=6.5$ meV compared with $E_r=7$ meV for BaFe$_{1.904}$Ni$_{0.096}$As$_2$.
Figure 13(c) shows a wave vector cut along the $[1,K]$ direction at $E=6.5\pm 1$ meV, which confirm the transverse incommensurate nature of the resonance.
A similar cut along the $[H,0]$ direction indicates that the mode is commensurate along the longitudinal direction [Fig. 13(d)].

\begin{figure*}[t]
\includegraphics[width=0.7\textwidth]{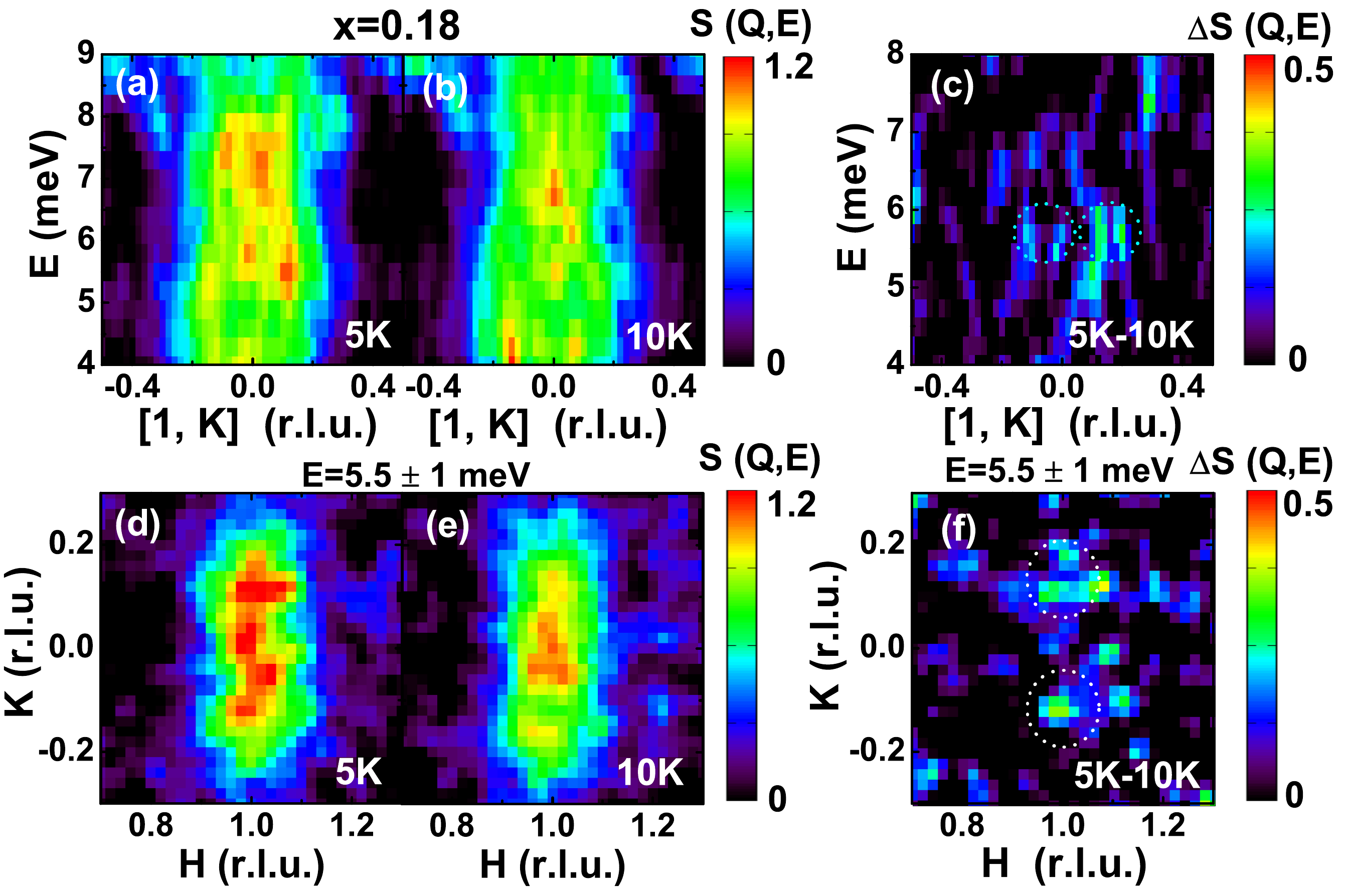}
\caption{
(Color online)
Temperature dependence of the spin excitations in BaFe$_{1.82}$Ni$_{0.18}$As$_{2}$ measured with $E_i=20$ meV.
(a,b,c) Energy dependence of the two-dimensional slices along the $Q=[1, K]$ direction at $T=5$ K, 10 K, and their differences, respectively.
(d,e,f) Wave vector dependence of the two-dimensional slices in the
energy range $E=5.5\pm 1$ meV at 5 K and 10 K, and their difference, respectively.  The dashed circles mark positions of incommensurate
spin fluctuations.
}
\end{figure*}

Turning our attention to a more electron overdoped sample BaFe$_{1.82}$Ni$_{0.18}$As$_{2}$ with $T_c=8$ K,
we were unable to find any magnetic signal in previous triple-axis measurements using 8 grams of sample \cite{hqluo12}.
Using 25 grams of co-aligned single crystals with an incident neutron beam energy of $E_i=20$ meV along the $c$-axis,
we can now detect clear low-energy spin excitations at the AF wave vector positions on MERLIN.
Figures 14(a) and 14(b) show spin excitation images projected onto the energy and $[1,K]$ plane at $T=T_c-3=5$ K and $T=T_c+2=10$ K, respectively.
Consistent with the behavior of spin excitations at other Ni-doping levels, we see plumes of scattering stemming from ${\bf Q}_{AF}=(1,0)$.  In the normal state (10 K),
spin excitations are commensurate and centered at ${\bf Q}_{AF}=(1,0)$ from $E=4$ meV to 9 meV [Fig. 14(b)].  On cooling to below $T_c$ (5 K), the scattering is enhanced between
$E=5$ meV and 7 meV [Fig. 14(a)].  The temperature difference plot in Figure 14(c) reveals evidence for incommensurate spin excitations.

Figures 14(d) and 14(e) show wave vector dependence of the spin excitations in the $[H,K]$ plane at
the resonance energy $E_r=5.5\pm 1$ meV below and above $T_c$, respectively.  In the normal state (10 K), spin excitations form transversely elongated ellipse commensurate with the underlying lattice [Fig. 14(e)].  On cooling
to below $T_c$ (5 K), spin excitations at transversely incommensurate positions are enhanced [Fig. 14(d)].   The temperature differences between 5 K and
10 K reveal transversely incommensurate spin excitations marked by dashed circles [Fig. 14(f)].

\begin{figure}[t]
\includegraphics[width=0.45\textwidth]{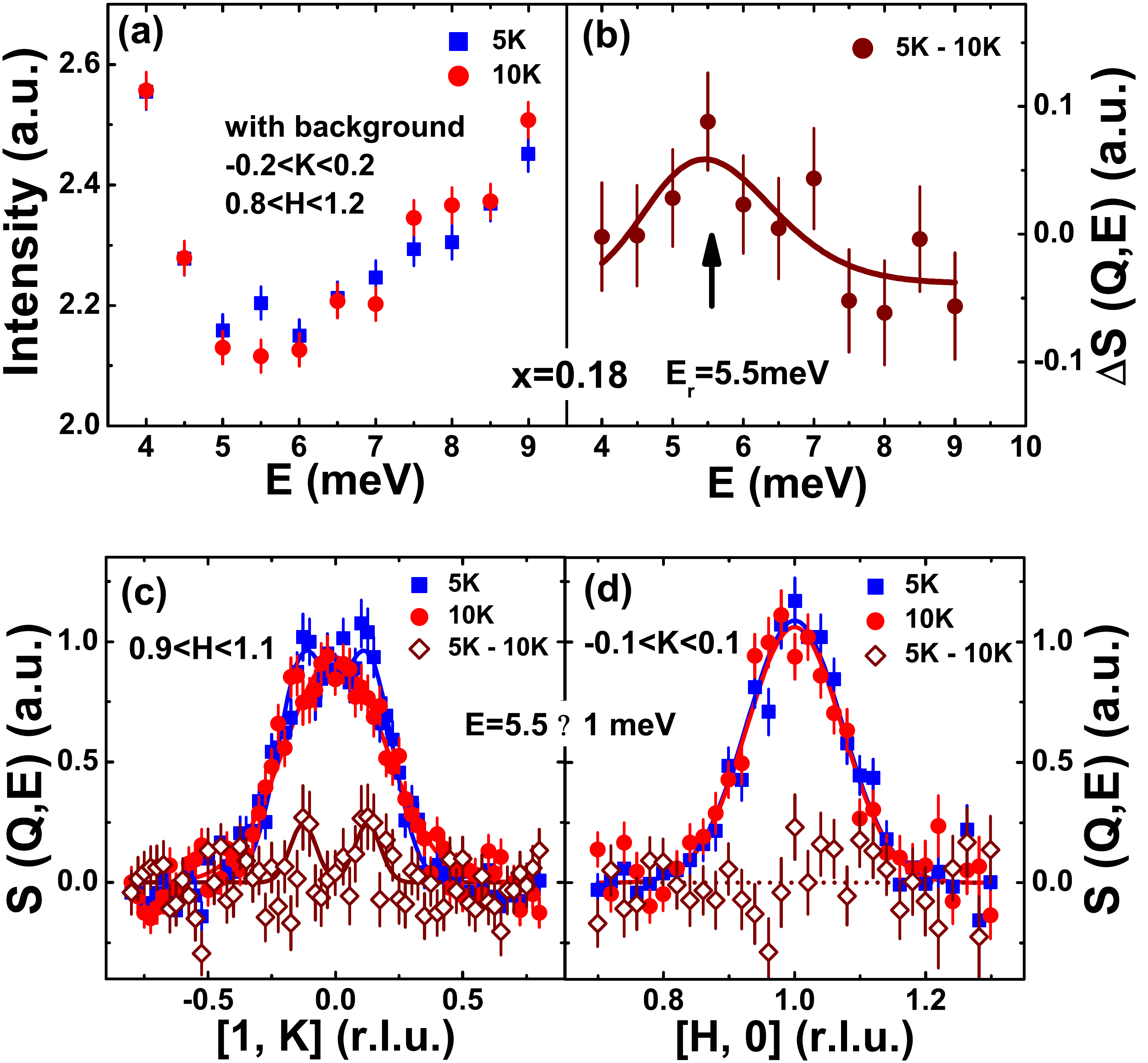}
\caption{(Color online)
(a,b) Energy dependence of the low energy spin excitations
in BaFe$_{1.82}$Ni$_{0.18}$As$_{2}$ at 5 K and 10 K, and their difference.
(c,d) Wave vector dependence of the low energy spin excitations in the energy range
$E=5.5\pm 1$ meV at 5 K and 10 K, and their difference.
}
\end{figure}

To further probe the wave vector, energy, and temperature dependence of the magnetic excitations in
BaFe$_{1.82}$Ni$_{0.18}$As$_{2}$, we show in Fig. 15(a) the energy dependence of the spin excitations near the AF ordering position integrated
within the range of $-0.2<K<0.2$ and $0.8<H<1.2$ r.l.u. below and above $T_c$.  The data reveals a small enhancement of the scattering below
$T_c$ for energies around $E_r=5.5$ meV.  Figure 15(b) shows the temperature difference between 5 K and 10 K, and one can see a very weak resonance
near $E_r=5.5$ meV.  Figure 15(c) shows cuts along the $[1,K]$ direction at $E=5.5\pm 1$ meV and $0.9<H<1.1$.  The red circles are data at 10 K showing a
commensurate peak centered at ${\bf Q}_{AF}=(1,0)$.  The blue squares are identical cut at 5 K, which have more scattering at the incommensurate positions.
The brown diamonds are the temperature difference plot which again reveal the incommensurate neutron spin resonance.  Figure 15(d) shows similar cuts along the
$[H,0]$ direction.  The scattering peaks at the commensurate AF ordering position and has
no observable changes across $T_c$, as confirmed by
the temperature difference plot shown as brown diamonds in Fig. 15(d).
Therefore, the resonance in electron-overdoped  BaFe$_{1.82}$Ni$_{0.18}$As$_{2}$ arises entirely from
superconductivity-induced incommensurate spin excitations.

\begin{figure}[t]
\includegraphics[width=0.4\textwidth]{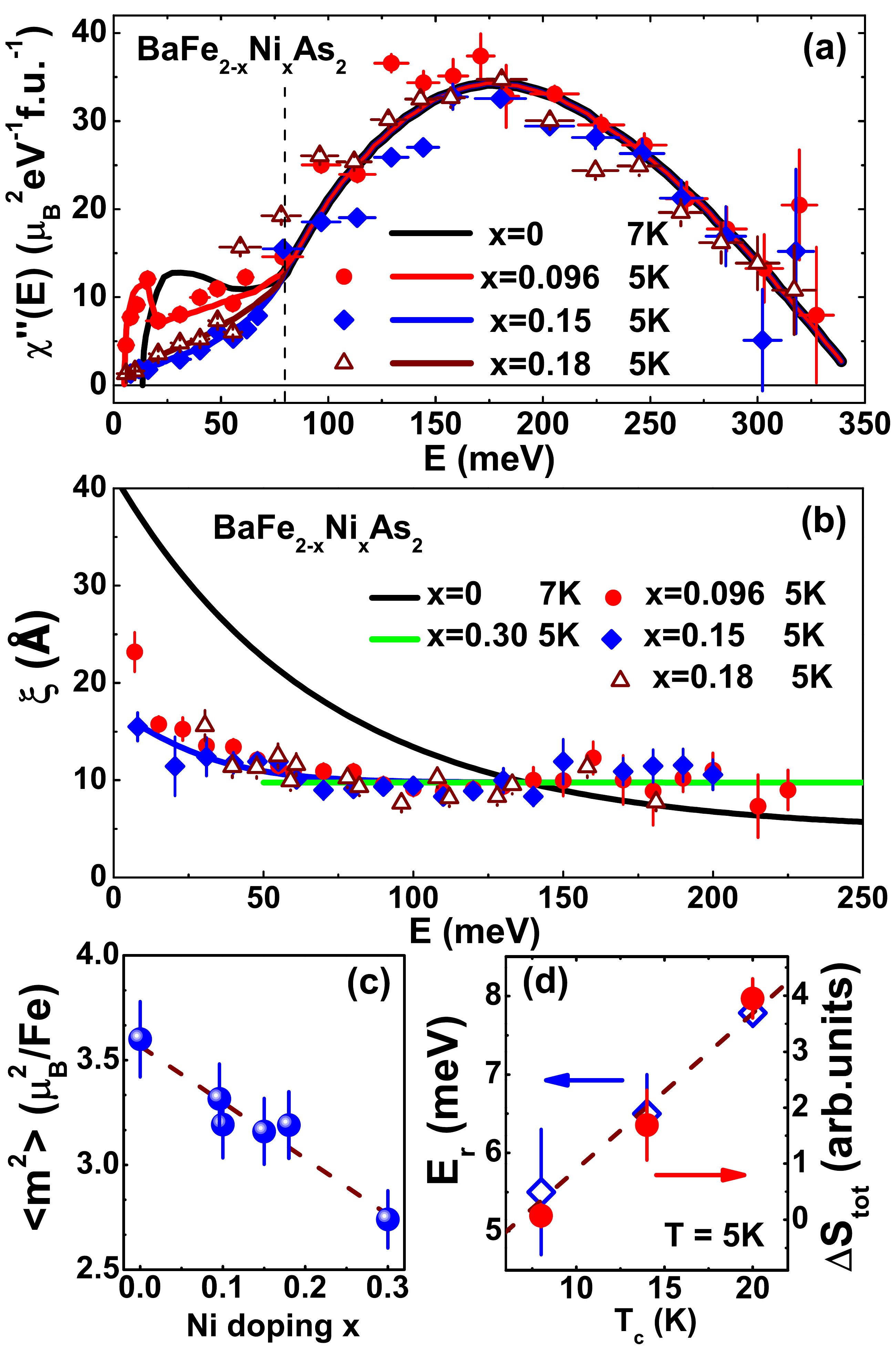}
\caption{(Color online)
(a) Energy dependence of the local dynamic susceptibility
$\chi^{\prime\prime}(E)$ for BaFe$_{2-x}$Ni$_{x}$As$_{2}$ with
$x=0,0.096,0.15,0.18$ in the absolute units ($\mu_B^2$/eV/f.u.).  While the high energy spin excitations are doping indepent,
low-energy spin excitations ($E<80$ meV) decreases with increasing electron doping.
(b) Energy dependence of the dynamic spin-spin correlation lengths ($\xi$) for
BaFe$_{2-x}$Ni$_{x}$As$_{2}$ with $x=0,0.096,0.15,0.18,0.30$ obtained by Fourier transform of the
constant-energy cuts along the $[1, K]$ direction. (c) Ni-doping dependence of the total fluctuating moment. (d)
The $T_c$ dependence of the resonance energy and its spectral weight.
}
\end{figure}

In previous work \cite{msliu12,mwang13}, we have established the electron doping evolution of the local dynamic susceptibility
$\chi^{\prime\prime}(E)$ for
BaFe$_{2-x}$Ni$_{x}$As$_{2}$ with $x=0,0.1,0.3$.  The new result serves to fill in the gap between the optimally electron-doped superconductor
and electron-overdoped nonsuperconductor.  Figure 16(a) shows the comparison of the energy dependent $\chi^{\prime\prime}(E)$ for $x=0,0.096,0.15$ and 0.18, using
method described before \cite{msliu12,mwang13}.
The solid line is the result for BaFe$_2$As$_2$ \cite{msliu12}.  The energy dependence of the local susceptibility for BaFe$_{1.904}$Ni$_{0.096}$As$_{2}$ is almost identical
to that of BaFe$_{1.9}$Ni$_{0.1}$As$_{2}$ \cite{msliu12}. On increasing the electron doping levels to $x=0.15$ and 0.18, we see a significant suppression of the local dynamic susceptibility for energies below $\sim$80 meV.  Instead of forming clear peak at the resonance energy as in the case of
BaFe$_{1.904}$Ni$_{0.096}$As$_{2}$, the energy dependent  $\chi^{\prime\prime}(E)$ increases linearly with increasing energy
and
the superconductivity-induced resonance is not a visible peak in BaFe$_{2-x}$Ni$_{x}$As$_{2}$ with $x=0.15$ and 0.18 [Fig. 16(a)].
For spin excitation energies above $\sim$80 meV, electron-doping to BaFe$_2$As$_2$ appears to have little effect on the local
dynamic susceptibility.  These results are consistent with the notion that
Fermi surface nesting and
itinerant electrons are controlling the low-energy spin excitations while high-energy spin excitations arise from the
local moments \cite{dai,khaule08,qmsi08,cfang08,ckxu08}.  Upon further doping to electron-overdoped nonsuperconductor for BaFe$_{2-x}$Ni$_{x}$As$_{2}$ with
$x>0.25$, Fermi surface nesting between the hole Fermi surface near $\Gamma$ and electron Fermi surface near $M$ point breaks down \cite{richard}, together with
vanishing superconductivity and low-energy spin excitations \cite{kmatan,flning10}.  However, high-energy spin excitations associated with local moments are not
affected \cite{mwang13}.  At present, it is unclear whether the large spin gap of $\sim$50 meV
in BaFe$_{1.7}$Ni$_{0.3}$As$_{2}$ \cite{mwang13} opens gradually or suddenly upon entering into the nonsuperconducting state with increasing
electron doping $x$.  Future work in this area might shed light on the relationship between the low-energy spin excitations and Fermi surface nesting.

Figure 16(b) shows the electron doping dependence of the dynamic spin-spin correlation lengths, obtained by Fourier transform of the $Q=[1, K]$ dependence of the spin
dynamic susceptibility \cite{msliu12}.  As we can see from the Figure, electron doping from an optimally doped superconductor to
electron-overdoped superconductor only appears to shorten the spin-spin correlation length for
spin excitations at low-energies, and have little impact to the zone boundary spin excitations.
To understand the impact of electron-doping
to the total fluctuating magnetic moments, defined as $\left\langle m^2\right\rangle=
(3/\pi)\int\chi^{\prime\prime}(E)dE/(1-\exp(-E/kT))$ \cite{clester10}, we show in Fig. 16(c) the electron-doping dependence of
$\left\langle m^2\right\rangle$ for BaFe$_{2-x}$Ni$_{x}$As$_{2}$ with $x=0,0.096,0.1,0.15,0.18,0.3$ \cite{msliu12,harriger12,mwang13}.  We used
$\left\langle m^2\right\rangle\approx 3.6\ \mu_B^2$/Fe for BaFe$_2$As$_2$ from a recent work \cite{harriger12}, a value slightly larger than
the earlier estimation of $\left\langle m^2\right\rangle\approx 3.17\pm0.16\ \mu_B^2$/Fe \cite{msliu12}.
The $\left\langle m^2\right\rangle$ shows a linear decrease in value
with increasing $x$.  From the electron doping dependence of the local dynamic susceptibility $\chi^{\prime\prime}(E)$ in Fig. 16(a), we see that
the decreasing total moment  $\left\langle m^2\right\rangle$ with increasing $x$
in BaFe$_{2-x}$Ni$_{x}$As$_{2}$
 is due almost entirely to the reduction in spin excitations below $\sim$80 meV.

Finally, Figure 16(d) shows the total spectral weight of spin resonance and the energy positions at $T=$ 5 K, estimated from the
superconductivity-induced spin excitation change, as a function of
$T_c$.  The resonance energy is linearly scaling with $T_c$, the same as previous results in
cuprates and pnictides \cite{mywang10,jtpark10,slli12}.  As superconductivity ceases to exist for BaFe$_{2-x}$Ni$_{x}$As$_{2}$ with $x\rightarrow 0.25$,
superconductivity-induced low-energy resonance also approaches zero, even though the high-energy spin excitations are not much affected.
This is consistent with the notion that superconductivity requires itinerant electron-spin excitation coupling \cite{mwang13}, and the
Fermi surface nesting driven low-energy
spin excitations are important for superconductivity in electron-doped iron pnictides.

\section{Discussion and Conclusions}

By comparing the structure, phase diagram, and magnetic excitations in
high-$T_c$ copper oxide, iron-based, and heavy Fermion
 superconductors, Scalapino concludes that spin fluctuation-mediated pairing is the
common thread linking different classes unconventional superconductors \cite{scalapino}.
Within the framework of this picture,
the superconducting condensation energy
should be accounted for by the change in magnetic exchange energy
$\Delta E_{ex}(T)$
between the normal ($N$) and superconducting ($S$)
phases at zero temperature.
For an isotropic $t$-$J$ model,
$\Delta E_{ex}(T)=2J[\left\langle {\bf S}_{i+x}\cdot{\bf S}_i\right\rangle_N-
\left\langle {\bf S}_{i+x}\cdot{\bf S}_i\right\rangle_S]$, where $J$ is the nearest neighbor magnetic exchange coupling and
$\left\langle {\bf S}_{i+x}\cdot{\bf S}_i\right\rangle$ is the magnetic scattering
in absolute units at temperature $T$ \cite{scalapino}.  If there are no changes in magnetic scattering
between the normal and superconducting state, spin excitations should not contribute to the superconducting condensation energy.
This is consistent with the observation that superconductivity-induced effect in spin excitations becomes very weaker in electron-overdoped
iron pnictides with reduced $T_c$.  While the total fluctuating moment $\left\langle m^2\right\rangle$ only decreases slightly on moving from
the AF parent compound BaFe$_2$As$_2$ to electron-overdoped nonsuperconducting BaFe$_{1.7}$Ni$_{0.3}$As$_{2}$ [Fig. 16(c)],
the changes in resonance intensity appears to correlate with superconducting $T_c$ [Fig. 16(d)].  This suggests that the superconducting transition
temperature in electron-doped iron pnictides is associated with the strength of the itinerant electron-low-energy spin excitations coupling or Fermi surface nesting
conditions of the hole and electron pockets.  This is not to say that high-energy spin excitations associated with local moments are not important for
superconductivity, as high energy spin excitations provide the basis for having a large effective magnetic exchange coupling $J$, which
is crucial for high-$T_c$ superconductivity \cite{mwang13}.

In conclusion, we use Triple-axis and TOF INS to study the temperature and electron-doping evolution of the spin excitations
in BaFe$_{2-x}$Ni$_{x}$As$_{2}$ with $x=0.096,0.15,0.18$.
Whereas the low-energy resonance induced by superconductivity becomes weak and vanishes near the electron
doping level when superconductivity ceases to exit, high-energy spin excitations are hardly modified by electron-doping and superconductivity.
For samples near optimal superconductivity, the FWHM of the resonance narrows in response to superconductivity.
We establish the dispersion of the resonance for $x=0.096$ sample near optimal superconductivity, and show that incommensurate spin excitations are prevalent in both the electron
underdoped and overdoped superconductors.  Although the total magnetic fluctuating moment only decreases slightly
with increasing electron-doping, the low-energy
spin excitations coupling with itinerant electrons vanishes when superconductivity is suppressed.  These results suggest that the Fermi surface nesting and low-energy spin excitation-itinerant electron coupling are
are critical for superconductivity in these materials.

\section{Acknowledgments}

The work at the
Institute of Physics, Chinese Academy of Sciences, is supported by the National Basic Research Program of China (No. 2011CBA00110 and 2012CB821400) and the National Science Foundation of China.
The work at Rice University is supported by the U.S. National Science Foundation through grant NSF-DMR-1063866
and NSF-OISE-0968226.

% Create the reference section using BibTeX:
%\bibliography{NoEndingPoint}

\end{document}